\documentclass[submission,Phys]{SciPost}

\pdfoutput=1
\usepackage{amsmath,amssymb}
\usepackage{booktabs,multirow,graphicx,tabularx,mathtools,slashed,url,hyperref}
\usepackage[maxfloats=100]{morefloats}
\usepackage{color,xcolor,braket,comment}
\usepackage[normalem]{ulem}
\usepackage{enumitem}
\usepackage{feynmp}

\makeatletter
\@ifundefined{pdfoutput}{}{\DeclareGraphicsRule{*}{mps}{*}{}}
\makeatother

\newcommand{\br}{\text{BR}}

\newcommand{\nz}[1]{\tilde{\chi}_{#1}^0}
\newcommand{\cp}[1]{\tilde{\chi}_{#1}^+}
\newcommand{\cm}[1]{\tilde{\chi}_{#1}^-}
\newcommand{\cpm}[1]{\tilde{\chi}_{#1}^\pm}

\newcommand{\mne}[1]{m_{\tilde{\chi}^0_{#1}}}

\newcommand{\qqquad}{\qquad \qquad}
\newcommand{\qqqquad}{\qquad \qquad \qquad}
\setlist[itemize]{itemsep=1pt,parsep=1pt, topsep=3pt}
\setlist[enumerate]{itemsep=1pt,parsep=1pt, topsep=3pt}

\newcommand{\lumi}{\mathcal{L}}

\newcommand{\eg}{e.\,g.\ }
\newcommand{\met}{\slashchar{E}_T}

\newcommand{\mev}{{\ensuremath \mathrm{MeV}}}
\newcommand{\gev}{{\ensuremath \mathrm{GeV}}}

\newcommand{\arxiv}[1]{\href{http://arxiv.org/abs/#1}{arXiv:#1}}

\AtBeginDocument{
  \heavyrulewidth=.08em
  \lightrulewidth=.05em
  \cmidrulewidth=.03em
  \belowrulesep=.65ex
  \belowbottomsep=0pt
  \aboverulesep=.4ex
  \abovetopsep=0pt
  \cmidrulesep=\doublerulesep
  \cmidrulekern=.5em
  \defaultaddspace= .5em
  \setlength{\tabcolsep}{0.5em}
}

\def\slashchar#1{\setbox0=\hbox{$#1$}           % set a box for #1
   \dimen0=\wd0                                 % and get its size
   \setbox1=\hbox{/} \dimen1=\wd1               % get size of /
   \ifdim\dimen0>\dimen1                        % #1 is bigger
      \rlap{\hbox to \dimen0{\hfil/\hfil}}      % so center / in box
      #1                                        % and print #1
   \else                                        % / is bigger
      \rlap{\hbox to \dimen1{\hfil$#1$\hfil}}   % so center #1
      /                                         % and print /
   \fi}

\begin{document}

\begin{fmffile}{feynman}

\begin{center}{\Large \textbf{
Actual Physics behind Mono-X
}}\end{center}

\begin{center}
Elias Bernreuther,
Jan Horak,
Tilman Plehn, and
Anja Butter
\end{center}

\begin{center}
Institut f\"ur Theoretische Physik, Universit\"at Heidelberg, Germany
\end{center}

\begin{center}
\today
\end{center}

\section*{Abstract}
{\bf Mono-X searches are standard dark matter search strategies at the
  LHC. First, we show how in the case of initial state radiation they
  essentially collapse to mono-jet searches. Second, we systematically
  study mono-X signatures from decays of heavier dark matter
  states. Direct detection constraints strongly limit our MSSM
  expectations, but largely vanish for mono-Z and mono-Higgs signals
  once we include light NMSSM mediators. Finally, the decay topology
  motivates mono-W-pair and mono-Higgs-pair searches, strengthening
  and complementing their mono-X counterparts.}

\bigskip
\bigskip
\bigskip

\vspace{10pt}
\noindent\rule{\textwidth}{1pt}
\tableofcontents\thispagestyle{fancy}
\noindent\rule{\textwidth}{1pt}
\vspace{10pt}

\newpage
%%%%%%%%%%%%%%%%%%%%%%%%%%%%%%%%%%%%%%%%%%%%%%%%%%%%%%%%%%%%
\section{Introduction}
\label{sec:intro}

Dark matter is one of the great puzzles in fundamental physics~\cite{planck}, with
overwhelming evidence for an explanation in terms of new particles in
the mass range between axions and primordial black holes.  As a
research field, it is driven by rapid experimental progress supporting
many different search strategies, making it likely that the current
and next generations of experiments will provide a definitive answer
to several of the fundamental concepts and models.

No matter what physics hypothesis we base our dark matter searches on,
we need to ensure that from a quantum field theory perspective the
model makes sense~\cite{dm_intro,dm_review,dm_lectures}. This applies
to an effective theory approach~\cite{dmeft1,dmeft2}, leading order
Feynman diagrams nowadays called simplified models~\cite{simplified},
and actual UV completions of the Standard
Model~\cite{simplified_complete,martin,sascha}. Among the candidates
for the latter, supersymmetry~\cite{dm_susy,relic_surface} still
stands out for two reasons: first, it offers a dark matter model as
part of a perturbative gauge theory which can be evolved to
fundamental scales; second, aside from a theoretical fine-tuning
argument hardly related to the dark matter sector, it does not lose its
attractive features once we include the current LHC
constraints~\cite{aachen}.\bigskip

Dark matter searches at colliders have a long history. Invisible
particles recoiling against visible particles have been searched for
at least since the UA1/UA2 days. The corresponding visible particles
can in principle be jets, leptons, photons, weak bosons, or Higgs
bosons. The corresponding searches have been dubbed mono-$X$ searches
and are often motivated through effective theory arguments. We take
the opposite approach and classify many of the viable mono-$X$
searches through two event topologies:
\begin{enumerate}
\item dark matter mediators in the hard process, recoiling against
  Standard Model particles from \textsl{initial state radiation};
\item production of heavier dark sector states followed by
  \textsl{dark matter decays} to Standard Model particles and the actual dark
  matter agent~\cite{nigel}.
\end{enumerate}
Dark matter combined with initial state radiation (ISR) of Standard Model
particles allows for a systematic comparison of different mono-$X$
channels~\cite{mono_x}. In Sec.~\ref{sec:isr} we will quantitatively
compare mono-jet~\cite{mono_j}, mono-photon~\cite{mono_a}, and
mono-$Z$~\cite{mono_z} production for an on-shell $Z'$-mediator at the
LHC.  Aside from the obvious question which mono-$X$ channel works
best, we will also ask what we learn from combining different such
mono-$X$ signatures.  Other mediators, for example including
(pseudo-)scalars coupling to gluons, prefer mono-jet signatures by
construction. Our findings can easily be generalized to a proper $2
\to 3$ process, except that in this case the LHC mono-$X$ rate will be
negligibly small.

The second topology is motivated for example by supersymmetric
electroweakinos. They describe dark matter as a combination of
singlet, doublet, and triplet representations of $SU(2)_L$ and
therefore include additional neutral and charged dark matter
particles. Decays of heavy neutralinos and charginos to the lightest
neutralino allow us to compare mono-$Z$~\cite{mono_z},
mono-$W$~\cite{mono_w}, mono-Higgs~\cite{mono_h} signatures in
Sec.~\ref{sec:fsr}.  A side aspect of this classification is that
invisible $Z$-decays and invisible Higgs decays are naturally included
in our approach. We decouple the heavy Higgs mediators, which are
already established as the motivation for mono-$Z$ and mono-Higgs
signatures~\cite{martin}.\bigskip

For a proper dark matter model, the combination of relic density and
direct detection constraints strongly cuts into the LHC signals,
especially for the mono-$Z$ case.  One way to avoid direct detection
constraints is to rely more on heavier neutralinos and charginos. This
leads us to consider mono-$W$-pair and mono-Higgs-pair signatures. 
We find that in contrast to the usual effective
theory scenarios, decay topologies prefer mono-$W$ over mono-$Z$
signatures, where the former include sizable contributions from
mono-$W$-pairs. The most flexible process in avoiding current
constraints is mono-Higgs-pairs, where already in the MSSM we can
largely decouple the LHC and direct detection processes.

Finally, in Sec.~\ref{sec:beyond} we expand the mediator sector and
study the singlet--singlino dark matter sector in the NMSSM. This effectively
decouples the relic density constraint from our LHC analysis. We focus
on the most constrained mono-$Z$ and the most flexible mono-Higgs-pair
signatures and show how the light scalar and pseudoscalar mediators
allow us to avoid the corner which the relic density and direct
detection constraints usually push us into. Instead, we find sizable
signal rates for the LHC in the presence of all available constraints.

%%%%%%%%%%%%%%%%%%%%%%%%%%%%%%%%%%%%%%%%%%%%%%%%%%%%%%%%%%%%
\section{Initial state X-radiation}
\label{sec:isr}

\begin{figure}[b!]
\begin{center}
\begin{fmfgraph*}(80,50)
\fmfset{arrow_len}{2mm}
\fmfleft{i1,i2}
\fmfright{o1,o2,o3}
\fmf{fermion,tension=0.8,width=0.6}{i1,v1}
\fmf{fermion,tension=0.6,width=0.6}{v1,v2}
\fmf{fermion,tension=0.4,width=0.6}{v2,i2}
\fmf{gluon,tension=0.4,width=0.6}{v1,o1}
\fmf{fermion,tension=0.5,width=0.6}{v2,o3}
\fmf{fermion,tension=0.7,width=0.6}{o2,v2}
\fmflabel{$q$}{i1}
\fmflabel{$\bar{q}$}{i2}
\fmflabel{$g$}{o1}
\fmflabel{$\chi$}{o2}
\fmflabel{$\chi$}{o3}
\fmfblob{.2w}{v2} 
\end{fmfgraph*}
\hspace*{0.1\textwidth}
\begin{fmfgraph*}(80,50)
\fmfset{arrow_len}{2mm}
\fmfleft{i1,i2}
\fmfright{o1,o2,o3}
\fmf{fermion,tension=0.8,width=0.6}{i1,v1}
\fmf{fermion,tension=0.6,width=0.6}{v1,v2}
\fmf{fermion,tension=0.4,width=0.6}{v2,i2}
\fmf{photon,tension=0.4,width=0.6}{v1,o1}
\fmf{fermion,tension=0.5,width=0.6}{v2,o3}
\fmf{fermion,tension=0.7,width=0.6}{o2,v2}
\fmflabel{$q$}{i1}
\fmflabel{$\bar{q}$}{i2}
\fmflabel{$\gamma$}{o1}
\fmflabel{$\chi$}{o2}
\fmflabel{$\chi$}{o3}
\fmfblob{.2w}{v2} 
\end{fmfgraph*}
\hspace*{0.1\textwidth}
\begin{fmfgraph*}(80,50)
\fmfset{arrow_len}{2mm}
\fmfleft{i1,i2}
\fmfright{o0,o1,o2,o3}
\fmf{fermion,tension=0.8,width=0.6}{i1,v1}
\fmf{fermion,tension=0.6,width=0.6}{v1,v2}
\fmf{fermion,tension=0.4,width=0.6}{v2,i2}
\fmf{photon,tension=0.4,width=0.6,label=$Z$}{v1,v3}
\fmf{fermion,tension=0.5,width=0.6}{v2,o3}
\fmf{fermion,tension=0.7,width=0.6}{o2,v2}
\fmf{fermion,tension=0.8,width=0.6}{o0,v3}
\fmf{fermion,tension=0.6,width=0.6}{v3,o1}
\fmflabel{$q$}{i1}
\fmflabel{$\bar{q}$}{i2}
\fmflabel{$\bar{f}$}{o0}
\fmflabel{$f$}{o1}
\fmflabel{$\chi$}{o2}
\fmflabel{$\chi$}{o3}
\fmfblob{.2w}{v2} 
\end{fmfgraph*}
\end{center}
  \caption{Feynman diagrams contributing to mono-$X$ production}
  \label{fig:feyn_monox}
\end{figure}
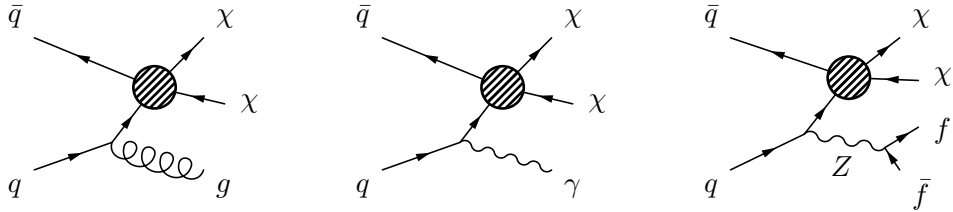
%------------------------------------------------------------

If we assume that the dark matter mediator couples to quarks, a
universal topology of dark matter signatures is given by initial state
radiation (ISR) of a gluon, a photon, or a $Z$-boson, shown in
Fig.~\ref{fig:feyn_monox}.  To illustrate their main features we
employ a model with a heavy $Z'$ mediator combined with a Majorana
fermion $\chi$ as a dark matter candidate. For our toy model the
mediator couples to the incoming quarks and to the dark matter
particles and can, if heavy, be integrated out. The signal process
then reads
\begin{align}
pp \to Z' X \to \chi \chi \; X 
\qqquad \text{with} \quad X=j,\gamma,Z \; .
\end{align}
For the mono-$Z$ signal we need to include a decay. While hadronic
decays $Z \to q\bar{q}$ come with a large branching ratio, leptonic
decays like $Z \to \ell \ell$ can help experimentally.  Mono-$W$
events can occur through ISR when we use a $q \bar{q}'$ initial state
to generate a hard $q\bar{q}$ scattering. Finally, mono-Higgs
signatures obviously make no sense when we rely on ISR.

Our toy model benefits from the phase space enhancement of an on-shell
$Z'$ mediator in the $s$-channel of the hard process, but the hard
process can be easily replaced by any other $s$-channel or $t$-channel
mediator exchange~\cite{dmeft2}. In that case the challenge will be to
enhance the cross section to explain the observed relic
density~\cite{planck} and predict a visible LHC signal for a
perturbative and predictive quantum theory. We only use our toy model
to illustrate the different mono-$X$ channels, for a discussion of
possible UV completions including a $Z'$ mediator we refer to the
detailed discussion in Ref.~\cite{sascha}.\bigskip

\begin{figure}[t]
\begin{center}
  \includegraphics[width=0.48\textwidth]{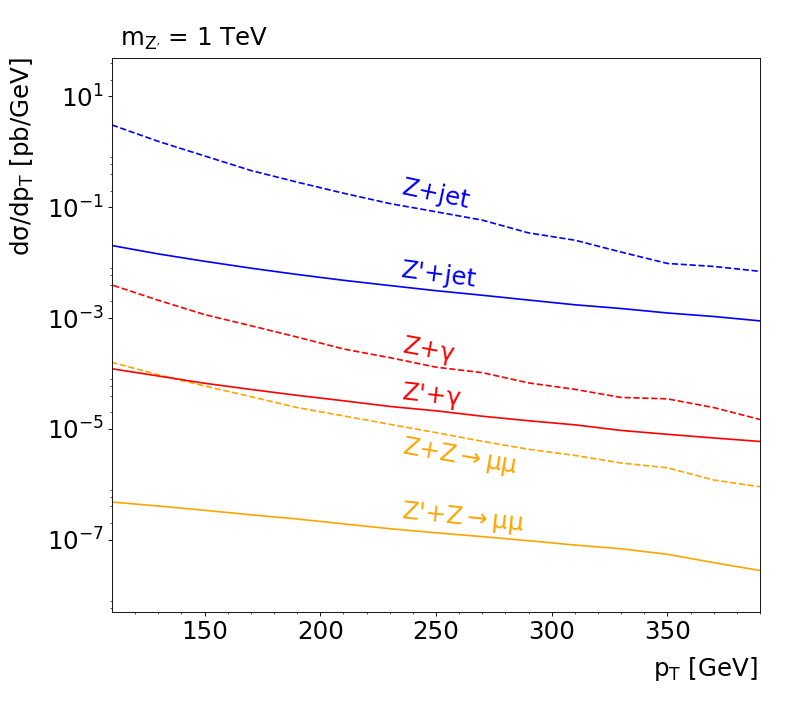}
  \hspace*{0.02\textwidth}
  \includegraphics[width=0.48\textwidth]{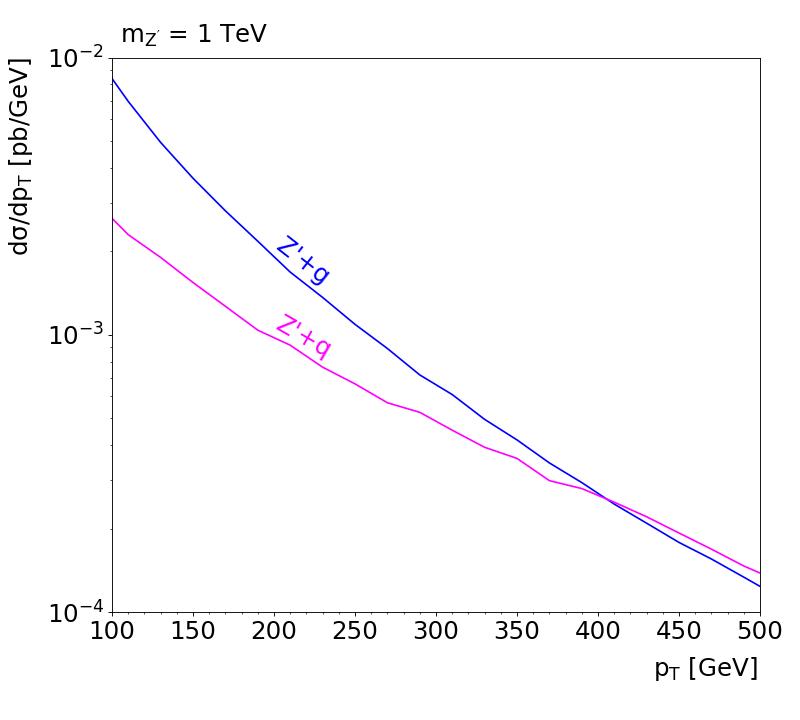}
\end{center}
  \caption{Transverse momentum spectrum for different mono-$X$ signals
    and backgrounds assuming a heavy vector mediator (left). The
    spectrum for the jet is composed by a $q\overline{q}$ initial
    state with a final state gluon and a quark gluon initial state
    with a quark jet radiated (right), here decomposed into fractions
    of the total signal.}
% The background is strongly dominated by quark jets in all momentum regimes.}
  \label{fig:jan}
\end{figure}

From the similarity of the three Feynman diagrams in
Fig.~\ref{fig:feyn_monox} we first derive that in the limit $m_Z \ll
m_{Z'}$ the total rates for the different mono-$X$ processes scale
like
\begin{align}
\frac{\sigma_{\chi \chi \gamma}}{\sigma_{\chi \chi j}} 
&\approx \frac{\alpha}{\alpha_s}  \frac{Q_q^2}{C_F}
\approx \frac{1}{40} \notag \\
\frac{\sigma_{\chi \chi \ell \ell}}{\sigma_{\chi \chi j}} 
&\approx \frac{\alpha}{\alpha_s} \frac{Q_q^2 s_w^2}{C_F} \; \br(Z \to \ell^+ \ell^-)
\approx \frac{1}{2000} \; .
\label{eq:monox_scaling}
\end{align}
Once we include the $Z$-mass the actual suppression of the mono-$Z$
channel is closer to $10^{-4}$.  In addition, the Feynman diagrams
also suggest that any kinematic $x$-distribution scales like
\begin{align}
\frac{1}{\sigma_{\chi \chi j}} \; \frac{d \sigma_{\chi \chi j}}{d x}
\approx \frac{1}{\sigma_{\chi \chi \gamma}} \; \frac{d \sigma_{\chi \chi \gamma}}{d x}
\approx \frac{1}{\sigma_{\chi \chi ff}} \; \frac{d \sigma_{\chi \chi ff}}{d x}
\; .
\end{align}
We show the $p_T$ distributions for the different mono-$X$ channels in
Fig.~\ref{fig:jan}, indicating that there is indeed no visible
difference between their shapes.  Once we include phase space information,
the suppression of the mono-photon becomes stronger, because the
rapidity coverage of the detector for jets extends to $|\eta| < 4.5$,
while photons rely on an efficient electromagnetic calorimeter with
$|\eta| <2.5$. On the other hand, photons can be detected to
significantly smaller transverse momenta than jets.

%------------------------------------------------------------
%------------------------------------------------------------

%------------------------------------------------------------
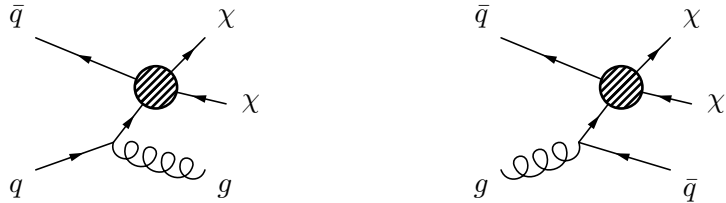
\begin{figure}[b!]
\begin{center}
\begin{fmfgraph*}(80,50)
\fmfset{arrow_len}{2mm}
\fmfleft{i1,i2}
\fmfright{o1,o2,o3}
\fmf{fermion,tension=0.8,width=0.6}{i1,v1}
\fmf{fermion,tension=0.6,width=0.6}{v1,v2}
\fmf{fermion,tension=0.4,width=0.6}{v2,i2}
\fmf{gluon,tension=0.4,width=0.6}{v1,o1}
\fmf{fermion,tension=0.5,width=0.6}{v2,o3}
\fmf{fermion,tension=0.7,width=0.6}{o2,v2}
\fmflabel{$q$}{i1}
\fmflabel{$\bar{q}$}{i2}
\fmflabel{$g$}{o1}
\fmflabel{$\chi$}{o2}
\fmflabel{$\chi$}{o3}
\fmfblob{.2w}{v2} 
\end{fmfgraph*}
\hspace*{0.2\textwidth}
\begin{fmfgraph*}(80,50)
\fmfset{arrow_len}{2mm}
\fmfleft{i1,i2}
\fmfright{o1,o2,o3}
\fmf{gluon,tension=0.8,width=0.6}{i1,v1}
\fmf{fermion,tension=0.6,width=0.6}{v1,v2}
\fmf{fermion,tension=0.4,width=0.6}{v2,i2}
\fmf{fermion,tension=0.4,width=0.6}{o1,v1}
\fmf{fermion,tension=0.5,width=0.6}{v2,o3}
\fmf{fermion,tension=0.7,width=0.6}{o2,v2}
\fmflabel{$g$}{i1}
\fmflabel{$\bar{q}$}{i2}
\fmflabel{$\bar{q}$}{o1}
\fmflabel{$\chi$}{o2}
\fmflabel{$\chi$}{o3}
\fmfblob{.2w}{v2} 
\end{fmfgraph*}
\end{center}
  \caption{Feynman diagrams contributing to mono-jet production.}
  \label{fig:feyn_monojet}
\end{figure}
%------------------------------------------------------------

Finally, the topology shown in Fig.~\ref{fig:feyn_monox} is not
complete for mono-jet production. In this case the gluon can be
crossed to the initial state, as shown in
Fig.~\ref{fig:feyn_monojet}. The size of this correction for dark
matter mediator radiation off quarks can be sizable and depends on
the transverse momentum of the jet, as shown in the right panel of
Fig.~\ref{fig:jan}.  On the other hand, the mediator could also couple
to incoming gluons and this way produce essentially only gluon
jets. Distinguishing these two mono-jet hypotheses is a perfect case
for including quark-gluon discrimination by default in any mono-jet
analysis.\bigskip

The same scaling arguments as for the mono-$X$ signal apply 
to the leading background,
\begin{align}
pp \to Z_{\nu \nu} X 
\qqquad \text{with} \quad X=j,\gamma,Z \; ,
\end{align}
possibly with the exception of mono-$Z$ production, where the hard
process and the collinear radiation are now both described by
$Z$-production. This means that the signal scaling of
Eq.\eqref{eq:monox_scaling} also applies to the leading backgrounds.
In the left panel of Fig.~\ref{fig:jan} we see that indeed all
mono-$X$ background scale similarly, and all background are slightly
softer than the corresponding signals. This is an effect of the
logarithmic collinear enhancement with the $Z'$ mass.

If our discovery channel is statistics limited, the significances
$n_\sigma$ for the different channels are given in terms of the
luminosity, efficiencies, and the cross sections
\begin{align}
n_{\sigma, j} 
=  \sqrt{\epsilon_j \lumi} \; 
  \frac{\sigma_{\chi \chi j} }{\sqrt{\sigma_{\nu \nu j}}} 
\qquad \Rightarrow \qquad 
n_{\sigma, \gamma} 
=&  \sqrt{\epsilon_\gamma \lumi} \; 
  \frac{\sigma_{\chi \chi \gamma} }{\sqrt{\sigma_{\nu \nu \gamma}}} 
\approx \frac{1}{6.3} \; \sqrt{ \frac{\epsilon_\gamma}{\epsilon_j}} \;
  n_{\sigma, j} \; .
\label{eq:monox_stat}
\end{align}
Unless the efficiency correction factors, including acceptance cuts
and cuts rejecting other backgrounds, point towards a very significant
advantage for the mono-photon channel, the mono-jet channel will be
the most promising search strategy. Using the same argument, the
expected mono-$Z$ significance will be negligible.  Note that the main
assumption behind this estimate is that the leading uncertainties are
statistical. For example with a jet energy scale uncertainty in ATLAS
around $1~...~2\%$, not far from the muon energy scale uncertainty,
this is definitely the case.\bigskip

An interesting aspect occurs when we include hadronic decays in
mono-$Z$ production~\cite{mono_zj}. In that case, the two jets neither
guarantee trigger, nor are they particularly useful to suppress
backgrounds. Instead, the boosted $Z$-boson contributes to the
mono-jet rate.  We can estimate the boost necessary to identify the
$Z$ as one jet from the separation of the two decay jets with the
momentum fractions $z$ and $1-z$,
\begin{align}
\Delta R_{jj} \approx \frac{m_Z}{p_{T,Z}} \; \frac{1}{\sqrt{z(1-z)}}
      > \frac{m_Z}{2 p_{T,Z}} \; .
\end{align}
This typical hyperbolic shape implies that $Z$-decays with for
instance $\Delta R_{jj} < 0.5$ are sensitive to events with $p_{T,Z} >
m_Z$. While this is an interesting observation, it does not help with
the ISR signature, because even including the hadronic $Z$-decays the
universal mono-jet rate above the same threshold is around 200 times
larger.\bigskip

Given the impressive control of ATLAS and CMS over their systematic
uncertainties it appears obvious, that ISR is not a valid
justification to search for dark matter production using the
mono-photon or mono-$Z$ signatures. Universally, mono-jet searches
will always be much more powerful. This is even more obvious when we
extend our models from the tree-level $Z'$ vector mediator to other
simplified or full models~\cite{dmeft2}. For example scalar mediators
in the $s$-channel preferably couple to gluons and clearly prefer
mono-jet searches; scalar, color-charged mediators in the $t$-channel
can always be produces on-shell and predict mono-jet events from ISR
as well as from on-shell production with a subsequent decay. The link
between mono-$X$ searches and such decays will be the subject of the
next section.

Not even the argument that we would like to study the properties of
dark matter by combining different LHC mono-$X$ channels holds in our
case. All we can learn from mono-photon and mono-$Z$ signals from ISR
topologies is collinear radiation of Standard Model particles off hard
incoming quarks.\footnote{While we are aware that this negative bottom
  line is known to many, we took the opportunity to illustrate it
  quantitatively as an introductory part of our study.}

%%%%%%%%%%%%%%%%%%%%%%%%%%%%%%%%%%%%%%%%%%%%%%%%%%%%%%%%%%%%
\section{Non-minimal dark matter sectors}
\label{sec:dm}

A much more promising topology leading to mono-$X$ signatures appears
for non-minimal dark matter sectors. They consist of a dark matter
agent, a SM or new physics mediator, and additional dark matter
states.  Such degrees of freedom arise when we embed dark matter in
$SU(2)_L$ multiplets.  The best-known example is the MSSM with mixed
bino, wino, and higgsino dark matter, plus a pair of Dirac charginos.
The electroweakino interactions reflect the supersymmetry of the
Lagrangian. Especially on the mediator side we gain additional freedom
from the NMSSM with its extended scalar sector. 

Note that this approach is exactly the opposite to the usual effective
field theory approaches, because it requires us to consider these
additional particles as propagating degrees of freedom at the LHC and
in the early universe.

%%%%%%%%%%%%%%%%%%%%%%%%%%%%%%%%%%%%%%%%%%%%%%%%%%%%%%%%%%%%
\subsection{MSSM}
\label{sec:dm_mssm}

\begin{figure}[t]
\begin{center}
\includegraphics[width=0.5\textwidth]{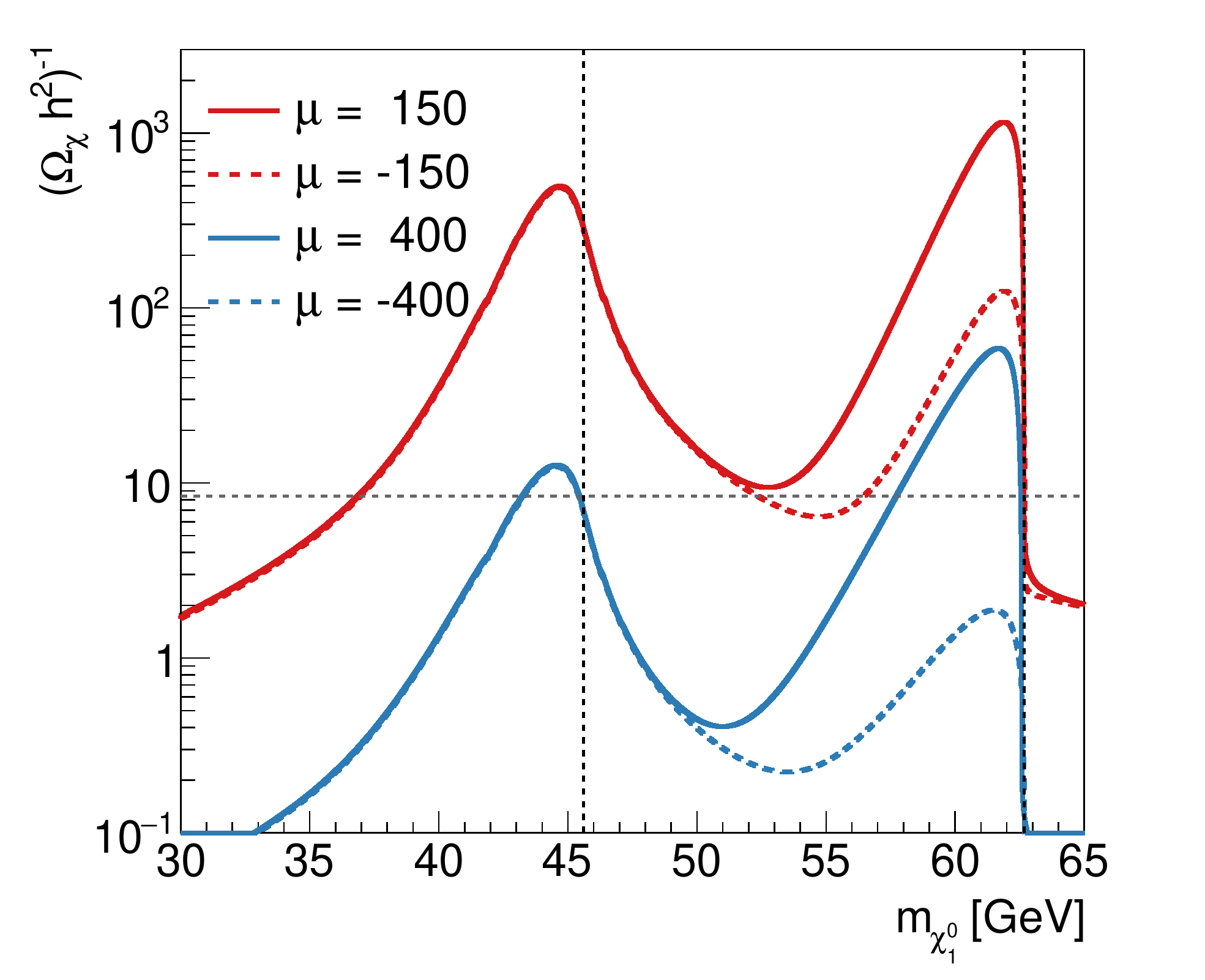}
\end{center}
\caption{Inverse relic density near the $Z$-pole and Higgs pole in the
  MSSM.}
\label{fig:higgsz_poles}
\end{figure}

The minimal supersymmetric electroweakino sector~\cite{dm_susy}
combines the fermionic partners of the weak gauge bosons and two Higgs
doublets. The corresponding mass matrix is
\begin{align}
%M_{\chi}=
\begin{pmatrix}
M_1 & 0 & -m_Z c_\beta s_w & m_Z s_\beta s_w \\
0 & M_2 & m_Z c_\beta c_w & -m_Z s_\beta c_w \\ 
-m_Z c_\beta s_w & m_Z c_\beta c_w & 0 & -\mu \\
m_Z s_\beta s_w & -m_Z s_\beta c_w & -\mu & 0 
\end{pmatrix}\; .
\label{eq:mssm_neut_matrix}
\end{align}
It is diagonalized through an orthogonal transformation $N$.  Two
higgsino doublets are not only required by the supersymmetric nature
of the MSSM Higgs sector, they also generally ensure that higgsino
loops do not lead to anomalies.  The annihilation process, which
guarantees the observed relic density in the MSSM, proceeds through a
set of mediators, namely the $Z$ and Higgs bosons of the Standard
Model, heavy new Higgs bosons, or new scalar partners of the Standard
Model fermions~\cite{mssm_ann}:
\begin{itemize}
\item $Z$-funnel annihilation through the higgsino components,
\begin{align}
g_{Z \nz{i} \nz{j}}
=\dfrac{g}{2 c_w}
 \left( N_{i3}N_{j3}-N_{i4}N_{j4}\right) \; .
\label{eq:z_coup}
\end{align}
  This coupling vanishes in the limit $t_\beta \rightarrow 1$ with equal higgsino
  fractions. Because the axial-vector component does not have a
  velocity suppression, the annihilation rate $\langle \sigma v
  \rangle$ prefers neutralino masses slightly above or below 45~GeV;
  directly on the $Z$-pole the annihilation is too efficient;
\item light $h$-funnel annihilation, where the dark matter mass is
  around $\mne{1} = m_h/2$~GeV, slightly away from the
  resonance. The underlying coupling
\begin{align}
g_{h \nz{i} \nz{j}} =
\frac{1}{2} \;
\left( g' N_{i1} - g N_{i2} \right) \; 
   \left( s_\alpha \; N_{j3} + c_\alpha \; N_{j4} \right) \;
+\left(i \leftrightarrow j\right)
\label{eq:h_coup}
\end{align}
  relies on higgsino-gaugino mixing. The angle $\alpha$ rotates the
  scalar Higgses into mass eigenstates. Given the SM-like nature of
  the light MSSM Higgs, almost the entire neutralino annihilation rate
  through the light Higgs funnel goes to $b\bar{b}$. The coupling then
  has the approximate form
\begin{align}
g_{h \nz{i} \nz{j}}
\approx
\frac{1}{2} \;
\left( g' N_{i1} - g N_{i2} \right) \; 
   s_\beta \left(- \frac{N_{j3}}{t_\beta} + N_{j4} \right) \; 
   + (i \leftrightarrow j)\;;
\label{eq:h_coup2}
\end{align}
\item heavy Higgs funnel annihilation, where the pseudoscalar $A$
  leads to efficient $s$-wave annihilation. The coupling is again
  driven by higgsino-gaugino mixing. Heavy scalar decays to down-type
  fermions are enhanced by $t_\beta$, which implies that for $t_\beta
  \gtrsim 30$ the resonance pole gets washed out and a $b\bar{b}$
  final state appears;
\item $t$-channel chargino exchange $\nz{1} \nz{1} \to WW$, relying on the coupling
\begin{align}
g_{W \nz{i} \cp{j}} =
g \; 
\left( \frac{1}{\sqrt{2}} N_{i4} V_{j2}^* - N_{i2} V_{j1}^* \right) \; ;
\label{eq:w_coup}
\end{align}
\item $t$-channel neutralino exchange, $\nz{1} \nz{1} \to ZZ$ or $\nz{1}
  \nz{1} \to hh$. For the annihilation into $Z$ pairs, the relevant axial-vector coupling is illustrated in
  Eq.\eqref{eq:z_coup}. For the annihilation to Higgs pairs, the
  relevant scalar coupling involves a product of higgsino and gaugino
  fractions, as given in Eq.\eqref{eq:h_coup};
\item $t$-channel sfermion exchange, \eg tau sleptons. In this case,
  significant coupling requires a large wino fraction, which typically
  leads to excessively large annihilation into $W$ bosons for dark
  matter masses below around 1~TeV;
\item co-annihilation channels are efficient whenever there is an
  additional supersymmetric particle within about 10\% of the dark
  matter
  mass~\cite{stau-co-annihilation,char-co-annihilation,stop-co-annihilation}.
  Additional light charginos or sfermions are strongly disfavored by
  LEP~\cite{lep_constraints}. Within the electroweakino sector,
  co-annihilation significantly contribute for example for
  processes with a light chargino in the $t$-channel.
\end{itemize}
In Fig.~\ref{fig:higgsz_poles} we illustrate the pole annihilation
through SM-like mediators in the MSSM for $M_1 = 10~...~80$~GeV, $\tan
\beta = 10$, and a decoupled wino. We see that because of the velocity
distribution the dark matter mass should actually be slightly below
the actual pole condition. While the velocity distribution is also
responsible for the width of the Higgs pole, the width of the $Z$-pole
is given by the physical $Z$-width. As discussed above, the neutralino
coupling to the $Z$-mediator only involves the higgsino fraction and
is therefore independent of the sign of $\mu$ or the exchange of the
two entries of $\mu$ in the neutralino mass matrix. On the other hand,
the gaugino-higgsino coupling to the SM-like Higgs is significantly
larger for $\mu >0$.\bigskip

%------------------------------------------------------------
%------------------------------------------------------------

On the side of the dark matter agent the MSSM ansatz is extremely
flexible, describing dark matter masses from tens of GeV to few
TeV~\cite{relic_surface}. For the three pure states, the neutralino
masses are fixed by the relic density prediction based on each
$SU(2)_L$ representation.  Wino dark matter, (co-)annihilating to
weak bosons has to be in the mass range of 2~TeV to 3~TeV. Higgsino
dark matter, annihilates slightly less efficiently, has to have a mass
between 1~TeV and 2~TeV.  Pure bino dark matter is only feasible with
very light sleptons in the $t$-channel.  One we include mixing, the
MSSM can explain the observed relic density over the entire mass
range.  The limitation of the MSSM is the number of mediators,
especially given the LHC constraints on heavy Higgs bosons and
$t$-channel sfermions. For dark matter at mass accessible to the LHC,
we essentially have to rely on the SM, or SM-like $Z$ and Higgs
mediators.

%%%%%%%%%%%%%%%%%%%%%%%%%%%%%%%%%%%%%%%%%%%%%%%%%%%%%%%%%%%%
\subsection{NMSSM}
\label{sec:dm_nmssm}

The NMSSM extension~\cite{nmssm_review,nmssm_ann} not only adds
another singlino to the neutralino sector, it also predicts a singlet
extension of the scalar mediator sector. The additional model
parameters in the scalar sector include
\begin{align}
\{ \, \lambda, A_\lambda, \kappa, A_\kappa
\}
\end{align}
with the convenient combinations 
\begin{alignat}{9}
\tilde \kappa &= \frac{\kappa}{\lambda}
&\qquad& \text{(singlino--higgsino mass ratio)} \notag \\
\tilde \lambda&= \frac{\lambda}{g} 
&\qquad& \text{(singlino--higgsino mixing)} \; .
\end{alignat}
The singlet mass entry in the extended
Higgs mass matrix is given by
\begin{align}
s_{2\beta} \dfrac{\tilde \lambda^2 A_\lambda}{2 \mu} 
  + \dfrac{\tilde \kappa  \mu}{m_Z^2} \left(A_\kappa+4\tilde \kappa \mu\right) \;
 .
\label{eq:singlet_mass}
\end{align}
The light scalar and pseudoscalar mediators from the new singlet
allow for very efficient dark matter annihilation and then predicts a
whole range of signatures at the LHC, including invisible Higgs
decays~\cite{nmssm_ann}. For our analysis we will set
\begin{align}
A_\lambda 
= 2 \mu \left( \dfrac{1}{s_{2 \beta}} - \tilde \kappa \right)
\label{eq:decoup_higgs}
\end{align}
at the relevant scale, to decouple the singlet sector from the SM-like
Higgs boson and avoid for example constraints from Higgs coupling
strengths.\bigskip

The neutralino in the NMSSM mass matrix has the same form as for the
MSSM, extended by an additional singlino
\begin{align}
\begin{pmatrix}
M_1 & 0 & -m_Z c_\beta s_w & m_Z s_\beta s_w & 0 \\
0 & M_2 & m_Z c_\beta c_w & -m_Z s_\beta c_w & 0 \\ 
-m_Z c_\beta s_w & m_Z c_\beta c_w & 0 & -\mu & -m_Z s_\beta \tilde \lambda \\
m_Z s_\beta s_w & -m_Z s_\beta c_w & -\mu & 0 & -m_Z c_\beta \tilde \lambda \\
0 & 0 & -m_Z s_\beta \tilde \lambda & -m_Z c_\beta \tilde \lambda & 2 \tilde \kappa \mu
\end{pmatrix} \; .
\label{eq:nmssm_neut_matrix}
\end{align}
The annihilation process to two SM fermions through a light scalar
or pseudo-scalar mediator
\begin{align}
\nz{1} \nz{1} \to a_s,h_s \to f \bar{f} 
\end{align}
is, in the limit of one SM-like Higgs boson, mediated by the coupling
\begin{align}
g_{a_s \nz{1} \nz{1}}
\approx g_{h_s \nz{1} \nz{1}}
\approx \lambda \sqrt{2}  \left(N_{13} N_{14}  - \tilde \kappa  N_{15}^2 \right) \; ,
\label{eq:nmssm_coupA_simp}
\end{align}
It allows for an efficient annihilation of much lighter dark matter
for a properly adjusted mediator mass. Going all the way to dark
matter with masses in the GeV range essentially avoids the Xenon-based
direct detection (DD) constraints and leaves us with CMB and
nucleosynthesis constraints.\bigskip

For extended dark matter sectors with charged particles a key
constraint comes from LEP. To be safe, we assume that any charged
particle which can hence be pair-produced in $e^+ e^-$ collisions and
which decays to leptons, jet, photons, or missing energy has to be
heavier than 103~GeV. This includes the light chargino, which through
its wino or higgsino mass parameters is closely tied to some of the
neutralinos.  Only the bino in the MSSM and the NMSSM, and the
singlino in the NMSSM can be lighter.

Finally, it is obviously possible to use mono-$X$ searches, or
following the previous discussion mono-jet searches, to target
electroweakinos. However, over most of the MSSM parameter space this
leads to proper $2 \to 3$ production processes, with the corresponding
phase-space suppression. The supersymmetric equivalent to the $2 \to
2$ topologies discussed in Sec.~\ref{sec:isr} would be searches for
invisible $Z$ or Higgs decays, which will be part of the discussion in
the following sections.

%%%%%%%%%%%%%%%%%%%%%%%%%%%%%%%%%%%%%%%%%%%%%%%%%%%%%%%%%%%%
\subsection{SFitter setup}
\label{sec:dm_sfitter}

%------------------------------------------------------------
\begin{table}[b!]
\begin{center}
\begin{tabular}{l l l}
\toprule
Observable & Constraint \\
\midrule
$\Gamma_{Z\to \chi \chi}$ & $<2$~MeV~\cite{lep_constraints} \\
$\Gamma_{h \to \chi \chi}$ & $<1.3$~MeV~\cite{hinv_ex} \\
%$\br_{h\to \text{inv}inv}$ & $< 0.24$ \\
$m_{\cpm{1}}$ & $>103.5$~GeV~\cite{lep_constraints}\\
$\Omega_\chi h^2$ & $0.1187 \pm 20\%$~\cite{planck} \\
$\sigma_{\mathrm{SI}}$ & Xenon1T \cite{xenon}, PandaX \cite{pandax} \\
$\sigma_{\mathrm{SD}}^p$ & Pico60 \cite{pico} \\
$\sigma_{\mathrm{SD}}^n$ & LUX \cite{lux} \\
\bottomrule
\end{tabular}
\end{center}
\caption{Overview of the constraints on the dark matter sector.}
\label{tab:constraints}
\end{table}
%------------------------------------------------------------

Our analysis of the MSSM and the NMSSM is based on the
\textsc{SFitter} framework~\cite{mssm_ann,nmssm_ann,sfitter} Because
we focus on the MSSM and NMSSM electroweakinos, we decouple all scalar
particles at 5~TeV, except for the SM-like Higgs and, in the NMSSM
case, the light set of scalar and pseudo-scalar mediators. This
includes the heavy 2HDM states, which can in principle play an
important role for dark matter annihilation~\cite{martin}.  The light
Higgs mass is adjusted to match the measured value $m_h = 125$~GeV
with the help of $\tan \beta \equiv t_\beta$ and $A_t$~\cite{m_h}.
All observables included in our global analysis are listed in
Tab.~\ref{tab:constraints}. We emphasize that for any well-defined
model the observed relic density is a crucial experimental
constraint. While searches for invisibly decaying particles at the LHC
certainly do not have to be related to this observable, any more
global interpretation in terms of dark matter will break down unless
we include a valid dark matter production mechanism and the relic
density constraint.

An interesting coincidence appears when we compare the invisible
Higgs~\cite{hinv_ex} and $Z$
decays~\cite{z_inv,lep_constraints}. While the branching ratio limits
are very different,
\begin{align}
\br_{Z\to \text{inv}} = 20.00\% \pm 0.06\%
\qqqquad 
\br_{h\to \text{inv}} < 24\% \; ,
\end{align}
the actual partial widths for a decay to dark matter are constrained
at very similar levels,
\begin{align}
\Gamma_{Z\to \chi \chi} < 2~\mev 
\qqqquad 
\Gamma_{h\to \chi \chi} < 1.3~\mev \; . 
\end{align}

Throughout our analysis we use the \textsc{SFitter} tool
box~\cite{sfitter}.  This includes calculating the particle spectrum
with \textsc{SuSpect3}~\cite{suspect}, the Higgs branching ratios with
\textsc{Susy-Hit} and \textsc{HDecay}~\cite{s-hit}, the relic density
and the direct detection rate with
\textsc{MicrOMEGAs}~\cite{micromegas}, and the LHC cross sections at
leading order with \textsc{Madgraph5}~\cite{madgraph5}.  Higher-order
corrections to the LHC cross sections are known~\cite{nlo_rates}, but
the NLO corrections are typically too small to make a difference to our
arguments.

%%%%%%%%%%%%%%%%%%%%%%%%%%%%%%%%%%%%%%%%%%%%%%%%%%%%%%%%%%%%
\section{Final state decays in the MSSM}
\label{sec:fsr}

To estimate the power of mono-$X$ analysis from final state decays we
need a dark matter model with several particles, where the heavier
states have an enhanced production rate at the LHC. Supersymmetric
winos and higgsinos are obvious and established candidates for such
searches. While for example the bino fraction allows us to explain the
relic density with a light neutralino, the winos and higgsinos couple
strongly to our SM mediators. We will discuss such signatures first
for the MSSM, where we have to negotiate a large LHC rate with the
relic density and direct detection (DD) constraints. Ignoring these
constraints would allow us to quote much large LHC rates, but we feel
that this would mean taking the experimentalists for a ride. Because
the main change in the NMSSM electroweakino sector is a new mediator,
we can use this extension to estimate an increased LHC reach from
non-SM mediators.

%%%%%%%%%%%%%%%%%%%%%%%%%%%%%%%%%%%%%%%%%%%%%%%%%%%%%%%%%%%%
\subsection{Mono-Z}
\label{sec:fsr_z}

%------------------------------------------------------------
\begin{figure}[b!]
\begin{center}
\begin{fmfgraph*}(80,50)
\fmfset{arrow_len}{2mm}
\fmfleft{i1,i2}
\fmflabel{$q$}{i1}
\fmflabel{$\bar{q}$}{i2}
\fmfright{o1,o2,o3}
\fmflabel{$\tilde{\chi}^0_1$}{o1}
\fmflabel{$\tilde{\chi}^0_1$}{o2}
\fmflabel{$Z$}{o3}
\fmf{fermion,width=0.6}{i1,v1,v3,i2}
\fmf{boson,width=0.6}{v3,o3}
\fmf{boson,label=$Z$,width=0.6}{v1,v2}
\fmf{boson,width=0.6}{o1,v2,o2}
\fmf{plain,width=0.6}{o1,v2,o2}
\end{fmfgraph*}
\hspace*{0.1\textwidth}
\begin{fmfgraph*}(80,50)
\fmfset{arrow_len}{2mm}
\fmfleft{i1,i2}
\fmflabel{$q$}{i1}
\fmflabel{$\bar{q}$}{i2}
\fmfright{o1,o2,o3}
\fmflabel{$\tilde{\chi}^0_1$}{o1}
\fmflabel{$\tilde{\chi}^0_1$}{o2}
\fmflabel{$Z$}{o3}
\fmf{fermion,width=0.6}{i1,v1,i2}
\fmf{boson,label=$Z$,width=0.6}{v1,v2}
\fmf{boson,width=0.6}{v2,o3}
\fmf{dashes,label=$h$,width=0.6}{v2,v3}
\fmf{boson,width=0.6}{o1,v3,o2}
\fmf{plain,width=0.6}{o1,v3,o2}
\end{fmfgraph*}
\hspace*{0.1\textwidth}
\begin{fmfgraph*}(80,50)
\fmfset{arrow_len}{2mm}
\fmfleft{i1,i2}
\fmflabel{$q$}{i1}
\fmflabel{$\bar{q}$}{i2}
\fmfright{o1,o2,o3}
\fmflabel{$\tilde{\chi}^0_1$}{o1}
\fmflabel{$\tilde{\chi}^0_1$}{o2}
\fmflabel{$Z$}{o3}
\fmf{fermion}{i1,v1,i2}
\fmf{boson,label=$Z$,width=0.6}{v1,v2}
\fmf{boson,width=0.6}{v2,o1}
\fmf{plain,width=0.6}{v2,o1}
\fmf{boson,width=0.6}{v2,v3}
\fmf{plain,label=$\tilde{\chi}^0_i$,width=0.6}{v2,v3}
\fmf{boson,width=0.6}{v3,o2}
\fmf{plain,width=0.6}{v3,o2}
\fmf{boson,width=0.6}{v3,o3}
\end{fmfgraph*}
\end{center}
  \caption{Feynman diagrams contributing to mono-$Z$ production in the
    MSSM, including initial-state $Z$-radiation with a $Z$-portal,
    $Zh$ production with a SM-like Higgs portal, and heavy neutralino
    decays.}
  \label{fig:feyn_monoz_mssm}
\end{figure}
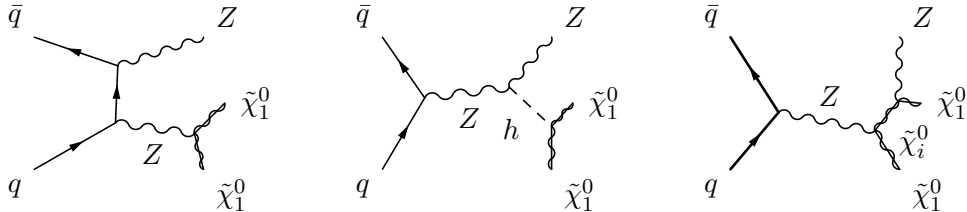
%------------------------------------------------------------

In the MSSM framework, mono-$Z$ production is defined as the hard process
\begin{align}
pp \to \nz{1} \nz{1} \; Z \; .
\label{eq:proc_monoz_mssm1} 
\end{align}
As long as we decouple the sfermions and heavy Higgs bosons, the
diagrams shown in Fig.~\ref{fig:feyn_monoz_mssm} are the only diagrams
contributing to this process at tree level. This means we can separate
three distinct topologies
\begin{alignat}{7}
pp &\to ZZ \to Z \; (\nz{1} \nz{1}) 
& \qqqquad &\text{ISR} \notag \\
pp &\to Zh \to Z \; (\nz{1} \nz{1}) 
& \qqqquad &\text{invisible Higgs decays} \notag \\
pp &\to \nz{j} \nz{1} \to (\nz{1} Z) \; \nz{1} 
& \qqqquad &\text{heavy neutralinos $j=2,3,4$} \; .
\end{alignat}
To avoid issues with gauge invariance we always include all topologies
in our simulation. If kinematically allowed, intermediate on-shell
states lead to a significant enhancement of the LHC production rate in
all three cases.

The first two topologies gain impact when the neutralinos are lighter
than 45~GeV or 62~GeV. Because of the LEP limits on charginos, this
implies that the dark matter agent cannot be a wino or a higgsino and
instead requires a sizable bino admixture. Based on the couplings
discussed in Sec.~\ref{sec:dm_mssm}, invisible $Z$-decays require a
large higgsino fraction, leading us to focus on bino--higgsino dark
matter. Similarly, invisible SM-like Higgs decays~\cite{hinv} require
gaugino--higgsino mixing, or in our case also bino--higgsino dark
matter.

For the third topology with its intermediate heavy neutralinos the
production process requires a sizable higgsino content in both of the
neutralinos involved. The decay $\nz{j} \to Z \nz{1}$ is mediated by the
same coupling, giving
\begin{align}
\sigma_{\nz{1} \nz{1} Z} \propto \frac{g_{Z \nz{1} \nz{j}}^4}{\Gamma_{\nz{j}}} \; .
\end{align}
It is then crucial that the mass difference between the two relevant
neutralinos is large, $\mne{j} - \mne{1} > m_Z$. For dominantly
higgsino dark matter with $m_Z \ll |\mu \pm M_1|, | \mu\pm M_2 |$ we
can approximate~\cite{masses}
\begin{align}
\mne{1,2} &= | \mu | +\frac{m_Z^2(1\pm s_{2\beta})(\mu \mp M_1c_w^2 \pm M_2s_w^2)}{2(\mu \mp M_1)(\mu \mp M_2)} \notag \\
%\mne{2} &=   | \mu | +\frac{m_Z^2(1- s_{2\beta})(\mu+M_1c_w^2+M_2s_w^2)}{2(\mu+M_1)(\mu+M_2)} \notag \\
\mne{2} - \mne{1} &= m_Z\left(\frac{m_Z}{M_2}c_w^2+\frac{m_Z}{M_1}s_w^2\right) \; .
\end{align}
This mass difference is always smaller than
$m_Z$~\cite{relic_surface,higgsinos}, again indicating that higgsinos
alone will not lead to a large mono-$Z$ signal. The obvious solution
is to again add a sizable bino content to the dark matter candidate
and analyze all three topologies in the limit
\begin{align}
M_1 < \lvert \mu \rvert \ll M_2 \; ,
\end{align}
with three propagating neutralinos.\bigskip

%------------------------------------------------------------
\begin{figure}[t]
\includegraphics[width=0.485\textwidth]{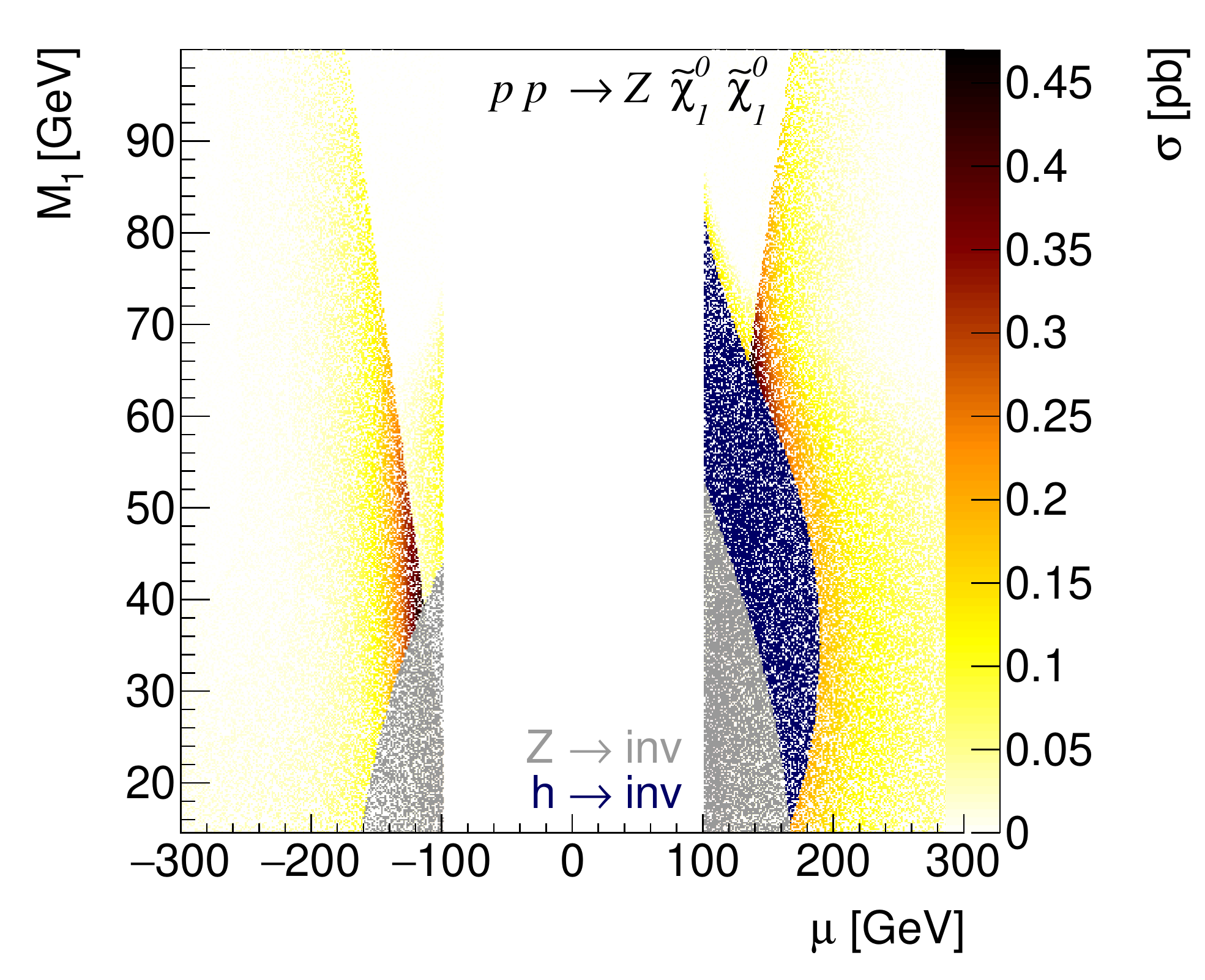}
\hspace*{0.02\textwidth}
\includegraphics[width=0.485\textwidth]{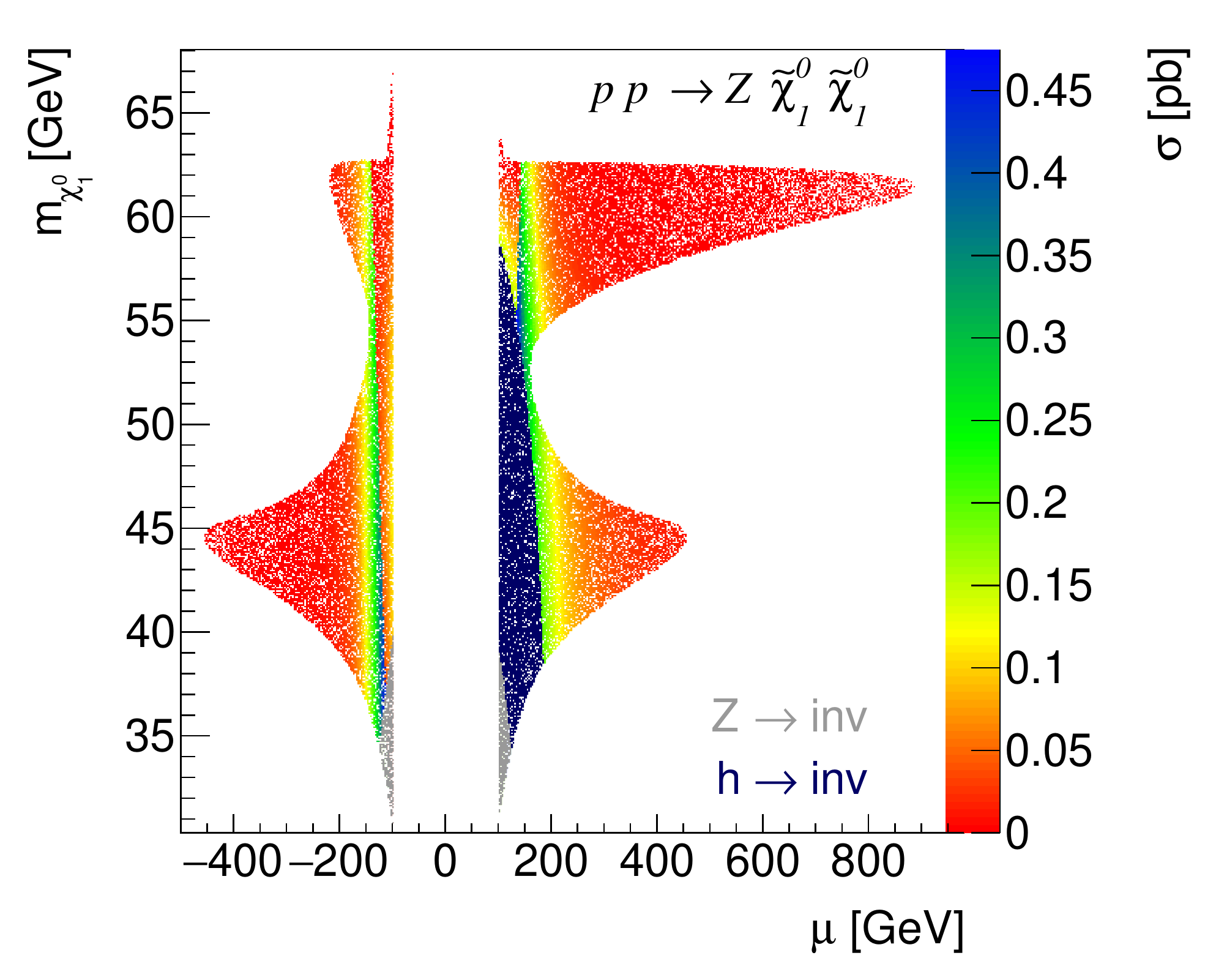}
   \caption{Cross section profiles for the mono-$Z$ process in the
     $\mu-M_1$ plane. All points fulfill the chargino mass bound
     (left) and, in addition, predict at most the measured relic
     density (right).  Regions excluded by invisible $Z$ and Higgs
     decays are shown in light gray and dark blue.}
   \label{fig:monozcs_lephinv}
\end{figure}
%------------------------------------------------------------

In the left panel of Fig.~\ref{fig:monozcs_lephinv} we show the
combined LHC production and decay rate for all three mono-$Z$
topologies in the $\mu-M_1$ plane. The dominant contribution to the
sizable rate slightly below the pb range comes from on-shell heavy
neutralinos. In the absence of all constraints, the slight asymmetry
in the sign of $\mu$ comes from the decay threshold as a function of
$\mu$ and $M_1$. Limits from invisible $Z$-decays constrain small
$M_1$ values through the dark matter mass and small $\lvert \mu
\rvert$ through the higgsino fraction. In contrast, invisible decays
of the SM-like Higgs require a large bino--higgsino mixing and are
therefore sensitive to the relative sign of $N_{13}$ and $N_{14}$ in
Eq.\eqref{eq:h_coup2}. This leads to a cancellation and hence weaker
constraints for $\mu<0$.\bigskip

As mentioned above, in any realistic thermal dark matter model the
observed relic density is a major constraint. Even the weaker
assumption that a given dark matter candidate only contributes a
fraction of the observed relic density translates into a relevant
lower limit on the dark matter annihilation rate. In the right panel
of Fig.~\ref{fig:monozcs_lephinv} we show the allowed parameter space
in terms of the dark matter mass $\mne{1}$ and $\mu$. The general
feature is that for a given dark matter mass the relic density defines
minimum coupling strengths for bino--higgsino dark matter, translated
into maximum values of $\mu$. The Higgs poles are highly asymmetric with respect to the sign of $\mu$, while the $Z$ poles are approximately symmetric. Because of the on-shell
enhancement of the annihilation rate, the invisible decay constraints
do not significantly constrain these parameter regions. Other
annihilation channels would appear for example for heavier dark
matter, but since we are interested in large LHC production rates we
limit ourselves to $\mne{1}< 70$~GeV at this stage.

%------------------------------------------------------------
\begin{figure}[t]
\includegraphics[width=0.485\textwidth]{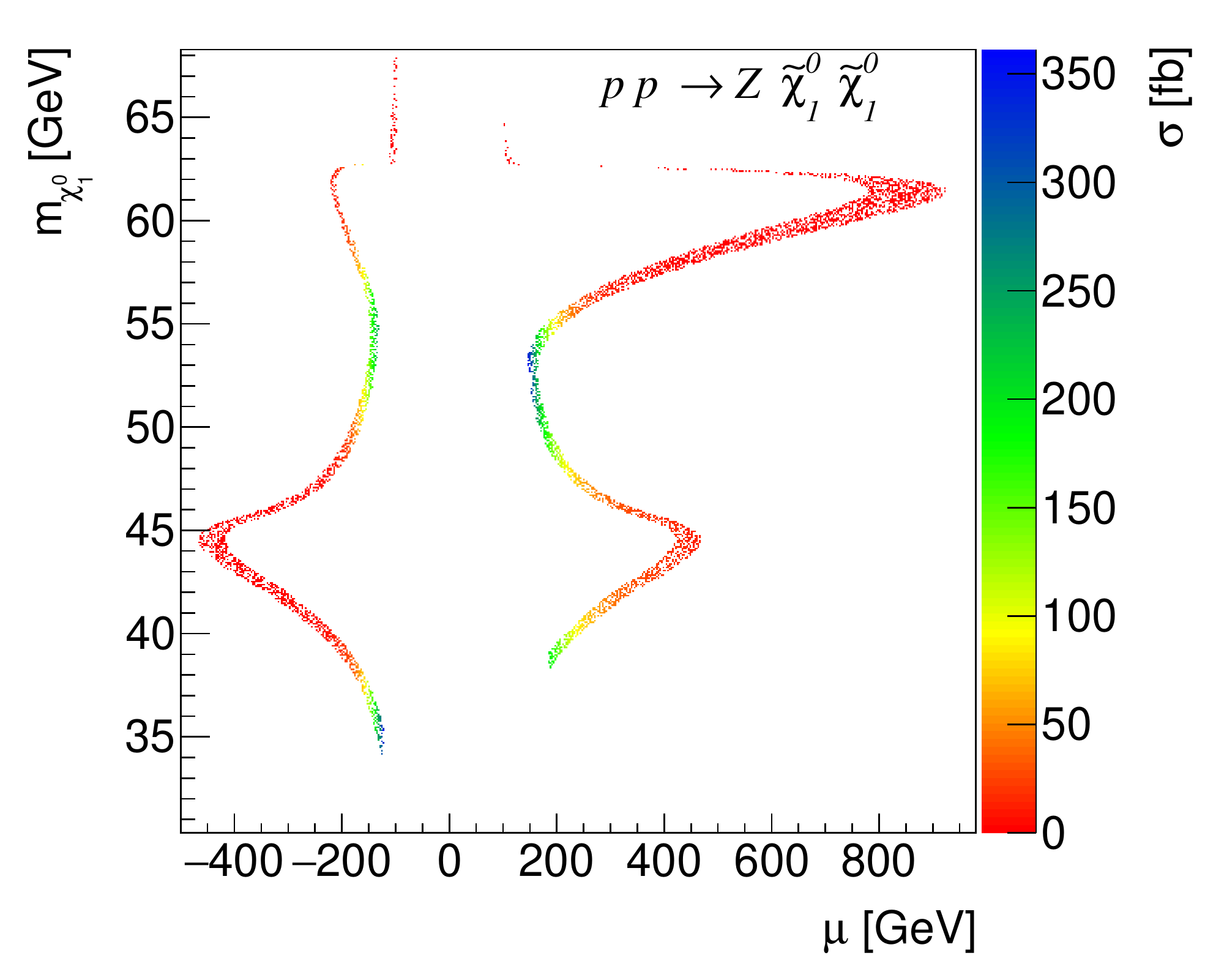}
\hspace*{0.02\textwidth}
\includegraphics[width=0.485\textwidth]{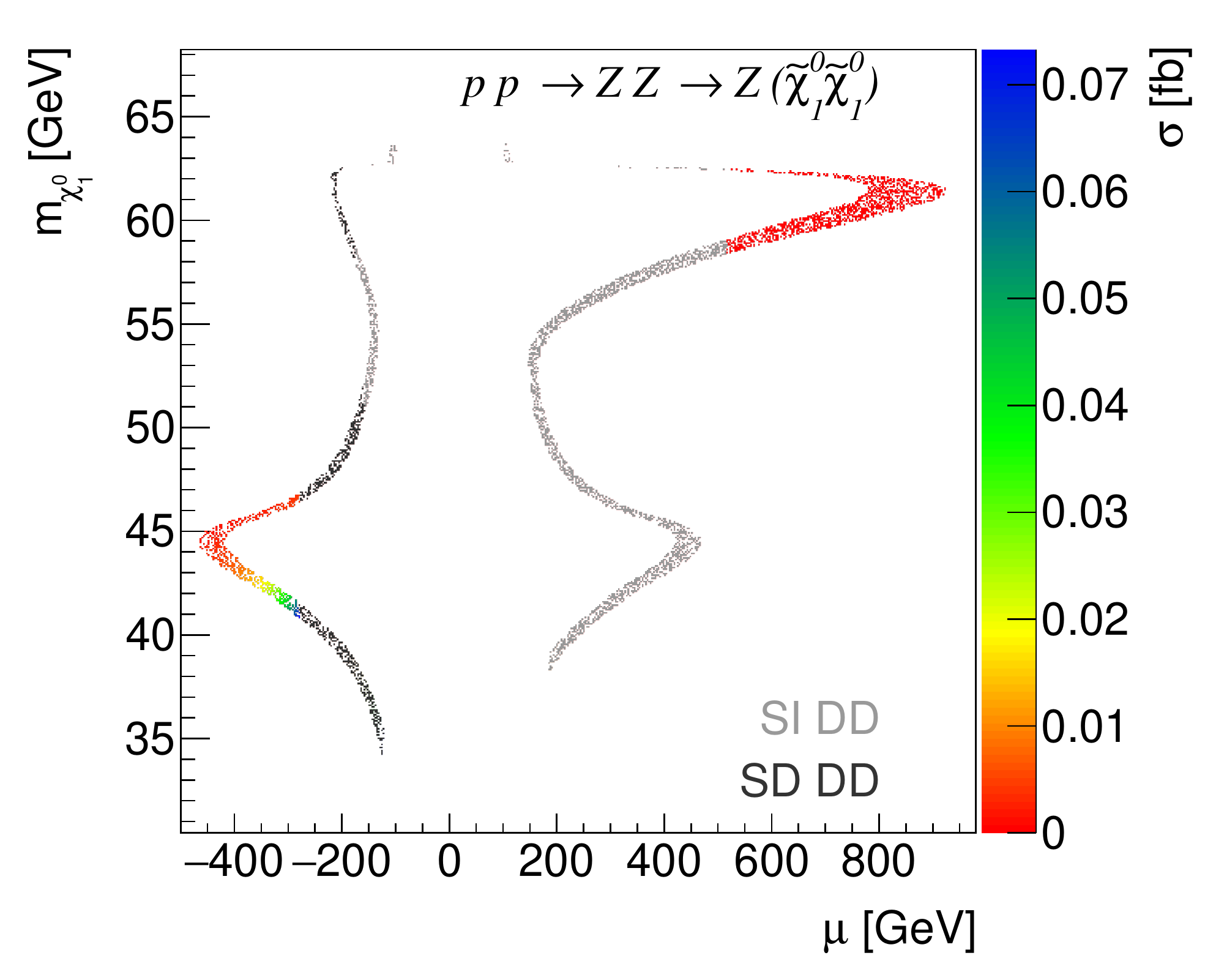} \\
\includegraphics[width=0.485\textwidth]{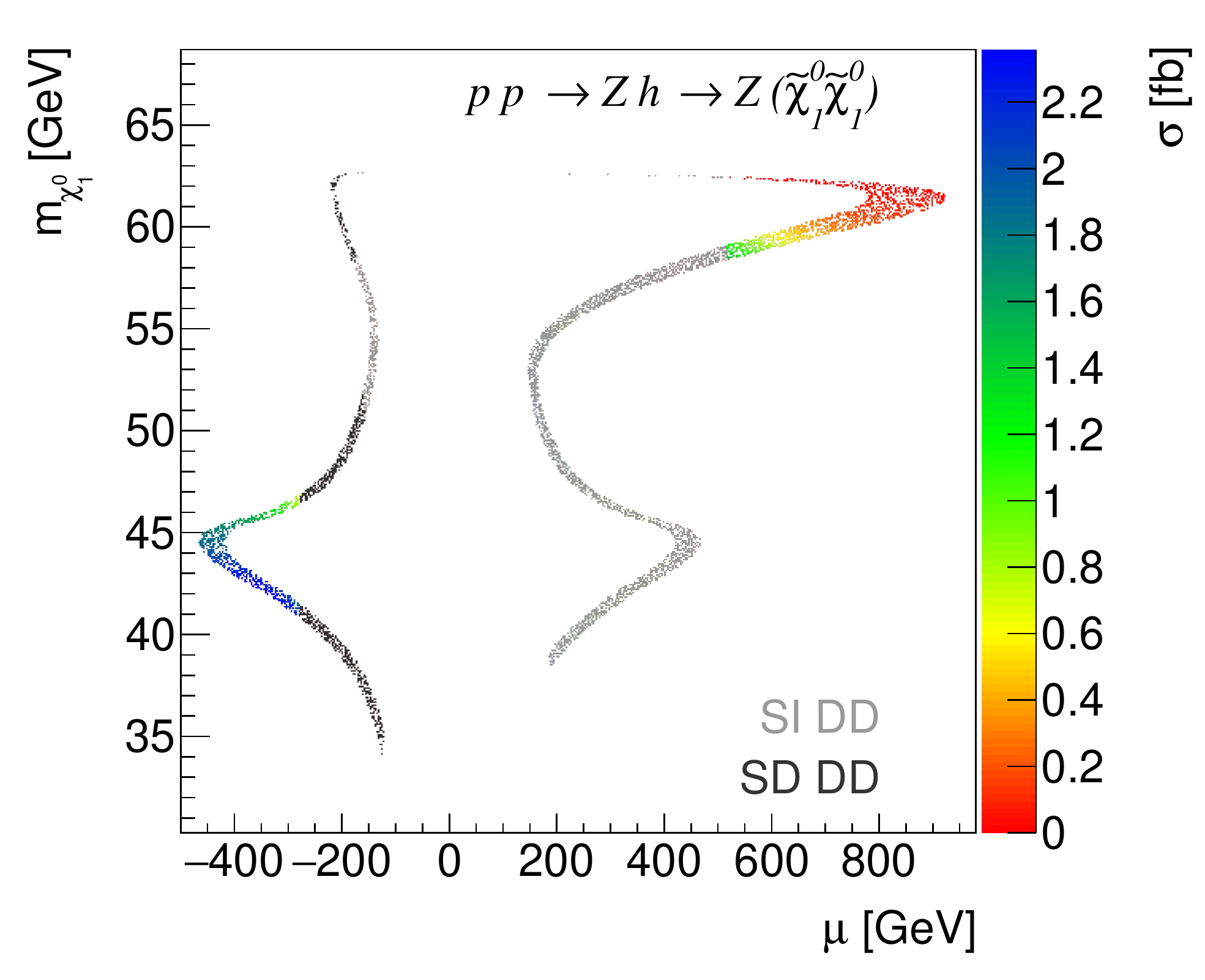} 
\hspace*{0.02\textwidth}
\includegraphics[width=0.485\textwidth]{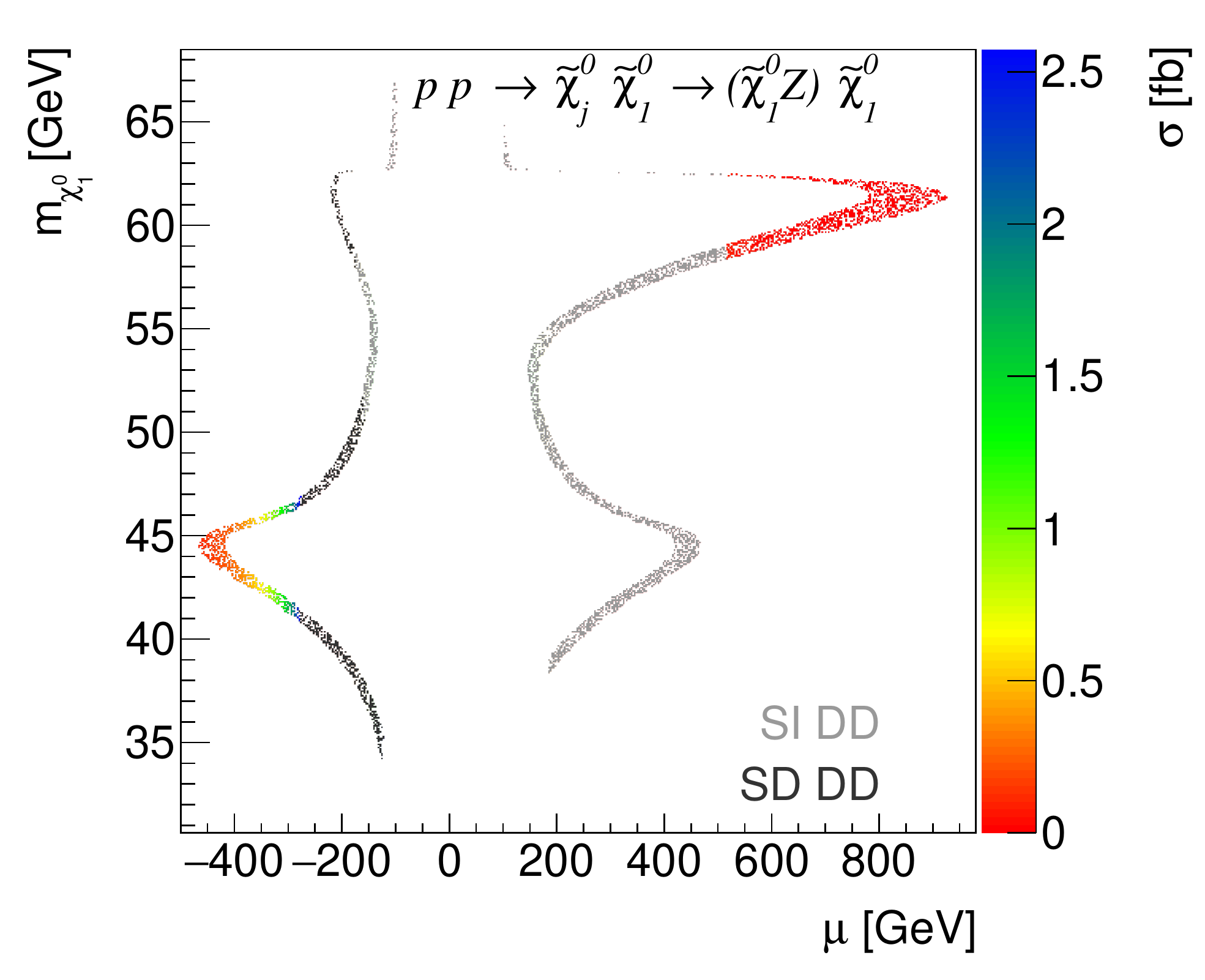}
   \caption{Mono-$Z$ cross sections in agreement with the observed
     relic density (upper left). For the three LHC topologies we also
     show the combination with DD limits (upper right to lower
     right). Points excluded by spin-independent DD limits are light
     gray, points excluded by spin-dependent direct detection in dark
     gray.}
   \label{fig:monozcs_relicdd}
\end{figure}
%------------------------------------------------------------

In the upper left panel of Fig.~\ref{fig:monozcs_relicdd} we start with all 
parameter points in agreement with the observed relic
density. The curve is identical to the shape shown in
Fig.~\ref{fig:monozcs_lephinv}. The important result is
that for the $Z$ and Higgs funnels the higgsino fractions are
relatively small, leading to mono-$Z$ rates around 10~fb at the
LHC. Larger LHC rates up to 350~fb are possible, but in regions where
the dark matter annihilation is not enhanced by on-shell diagrams.

When we include the exact relic density constraint, 
we should also consider the DD limits displayed in 
Tab.~\ref{tab:constraints}.  Based on the
spin-independent and spin-dependent interpretations the limits
translate into limits on the $g_{Z \nz{1} \nz{1}}$ and $g_{h\nz{1} \nz{1}}$
couplings, competitive with the full range of the on-shell peaks in
Fig.~\ref{fig:monozcs_lephinv}. In the remaining three panels of
Fig.~\ref{fig:monozcs_relicdd} we show all parameter points predicting
the observed relic density and indicate if they agree with the current
DD constraints.

The upper right panel of Fig.~\ref{fig:monozcs_relicdd} shows the
results for the ISR topology.  First, we observe some general features
from the interplay of the relic density constraint with
spin-independent and spin-dependent direct detection. Just like the
shape of the Higgs pole annihilation, the spin-independent constraints
are very asymmetric in the sign of $\mu$. This reflects the mixed
bino--higgsino coupling to the Higgs with a relative sign between
$N_{13}$ and $N_{14}$. Large preferred values of $\mu>0$ imply small
$g_{h\nz{1} \nz{1}}$ and correspond to the usual peak in the allowed
parameter space. This peak is not (yet) ruled out by direct
detection. For $\mu <0$ the spin-independent DD constraints are weak,
so the leading constraints are spin-dependent limits. Even for
$\mne{1} \approx m_h/2$ they are driven by $g_{Z \nz{1} \nz{1}}$.

As expected from our general ISR discussion in Sec.~\ref{sec:isr},
the expected LHC mono-$Z$ rates are very small. They reach 0.07~fb at
most, and in a very small region of parameter space around $\mne{1}
\approx 42$~GeV. This is the only region of parameter space where the
LHC process is still enhanced by an on-shell $Z$-decay, but the
couplings are not ruled out spin-dependent direct detection.

%-------------------------
\begin{figure}[t]
\begin{center}
\includegraphics[width=0.40\textwidth]{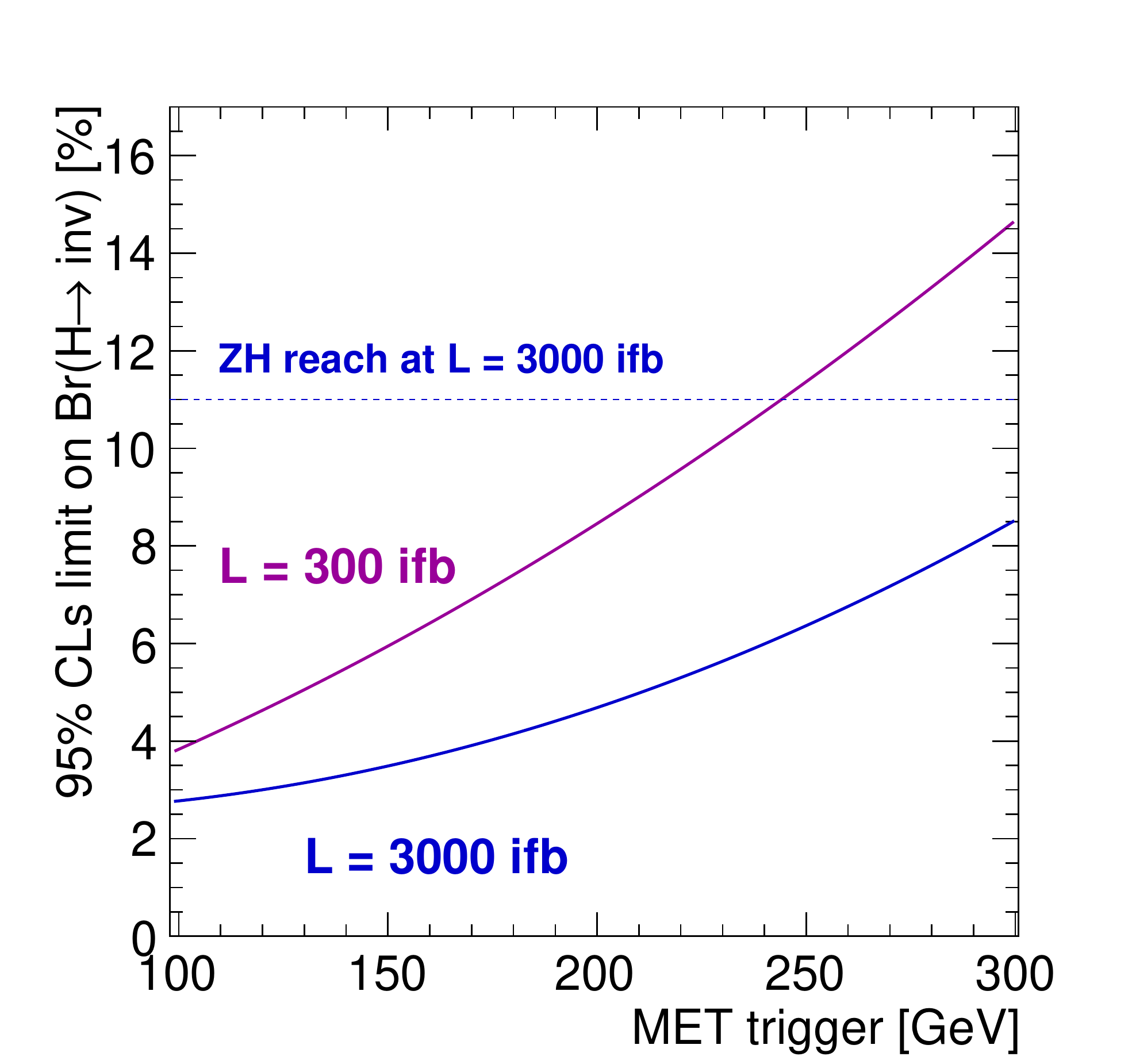}
\hspace*{0.1\textwidth}
\includegraphics[width=0.40\textwidth]{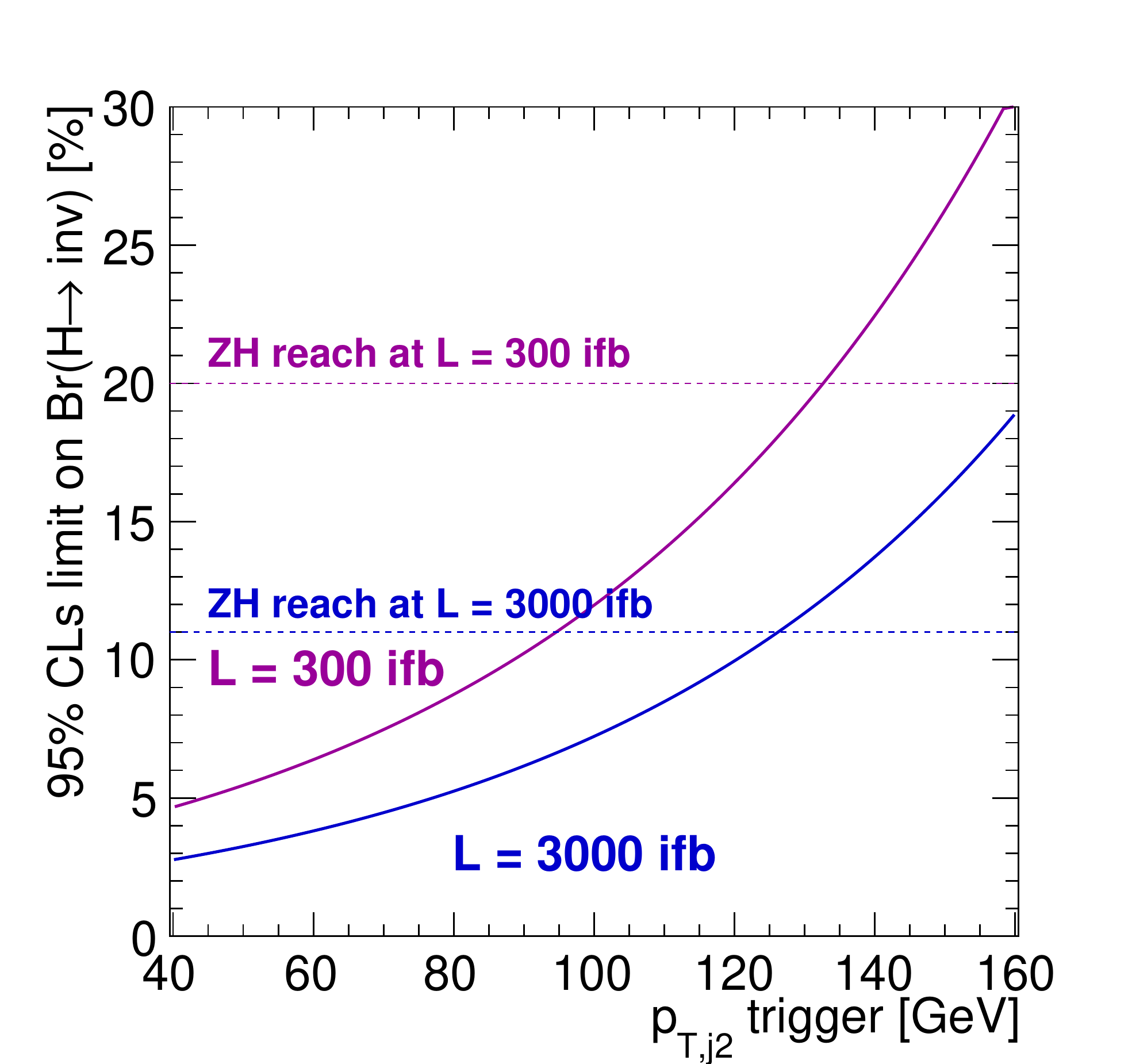}
\end{center}
\caption{CLs limits on invisible Higgs decays from weak boson fusion,
  as a function of trigger cuts on missing transverse energy (left) or
  the transverse momentum of the tagging jets (right), compared to the
  expected reach in the leptonic $Zh$ channel. Figures from
  Ref~\cite{anke_jennie}.}
\label{fig:anke_jennie}
\end{figure}
%-------------------------

The next, lower left panel shows the same information for the $Zh$
topology combined with invisible Higgs decays. The structure is
similar to ISR case, but with significantly large cross sections.  The
reason are the limits from invisible $Z$ and Higgs decays, which
following Sec.~\ref{sec:dm_sfitter} look similar in terms of the
partial width, but very different in terms of invisible branching
ratios. The latter are relevant for the different $(2\to2)$ mono-$Z$
channels.  Driven by the relic density constraint the largest rate for
the $Zh$ topology of around 2~fb appears for $\mne{1} \approx
41$~GeV. The large Higgs couplings are barely allowed by DD constraints. 

We can skip a dedicated analysis of mono-$Z$ production in the $Zh$
topology and instead resort to the literature~\cite{vh_inv}: the
problem is that we can search for exactly the same model using
invisible Higgs decays in weak boson
fusion~\cite{dieter,anke_jennie}. In Fig.~\ref{fig:anke_jennie} we
show the results from Ref.~\cite{anke_jennie} which indicate that even
with conservative assumptions on triggering at the high-luminosity LHC
the $Zh$ topology will never be the discovery channel for such dark
matter models.

Finally, we show the expected rates for on-shell neutralinos in the
lower right panel. Typical mono-$Z$ rates can reach 2.5~fb for light
dark matter, $\mne{1} = 40~...~47$~GeV. This window is given by the
relic density requirement, where annihilation off the $Z$-pole is
preferred because of the larger corresponding couplings. While the
rate for this topology does not reach the invisible Higgs rates, this
channel generally extends to larger dark matter masses. The limiting
factor is the lower limits on the heavier two higgsino masses and the
corresponding LHC production cross sections through an $s$-channel
$Z$.\bigskip

%------------------------------------------------------------
\begin{figure}[t]
	\includegraphics[width=0.485\textwidth]{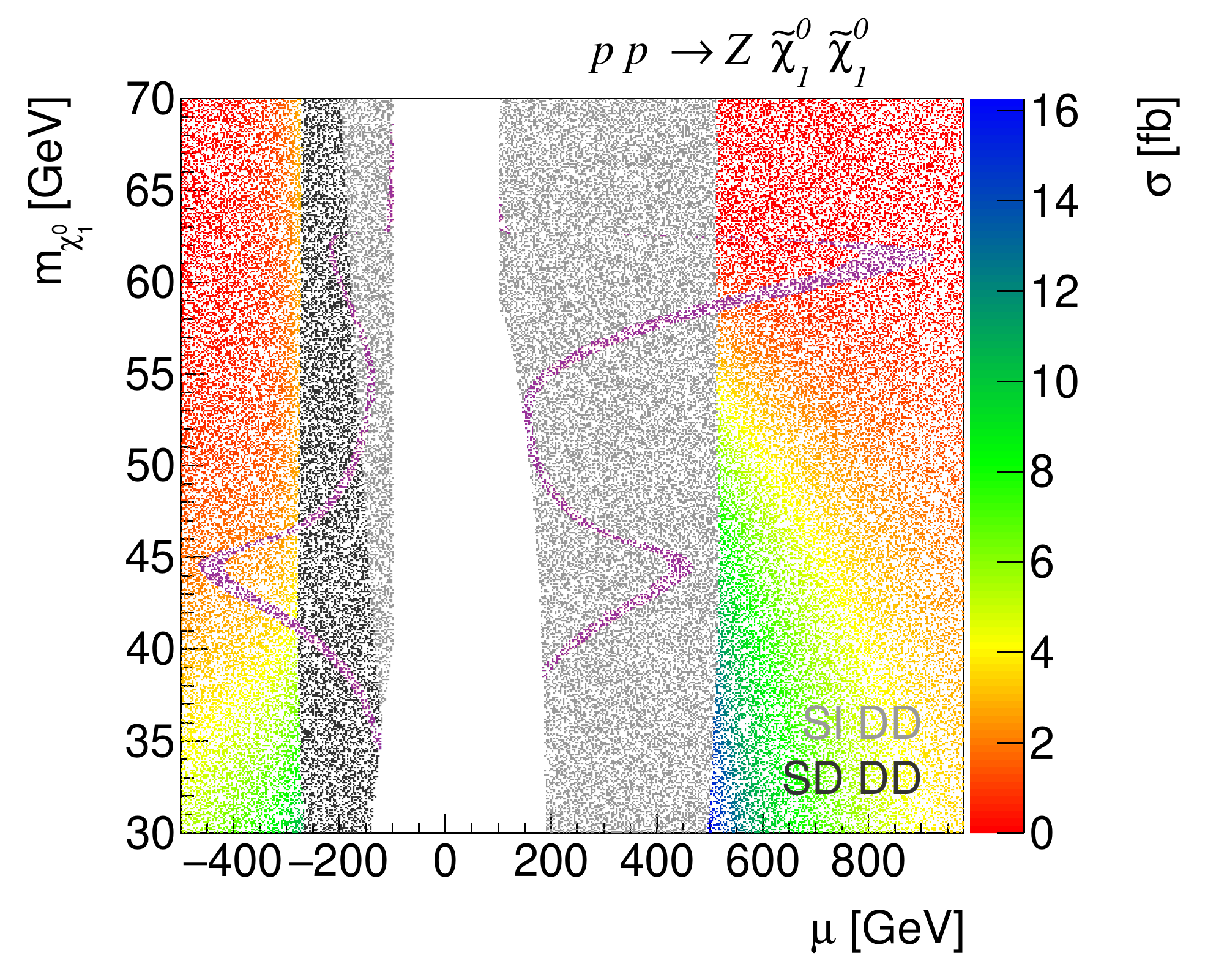}
	\hspace*{0.02\textwidth}
	\includegraphics[width=0.485\textwidth]{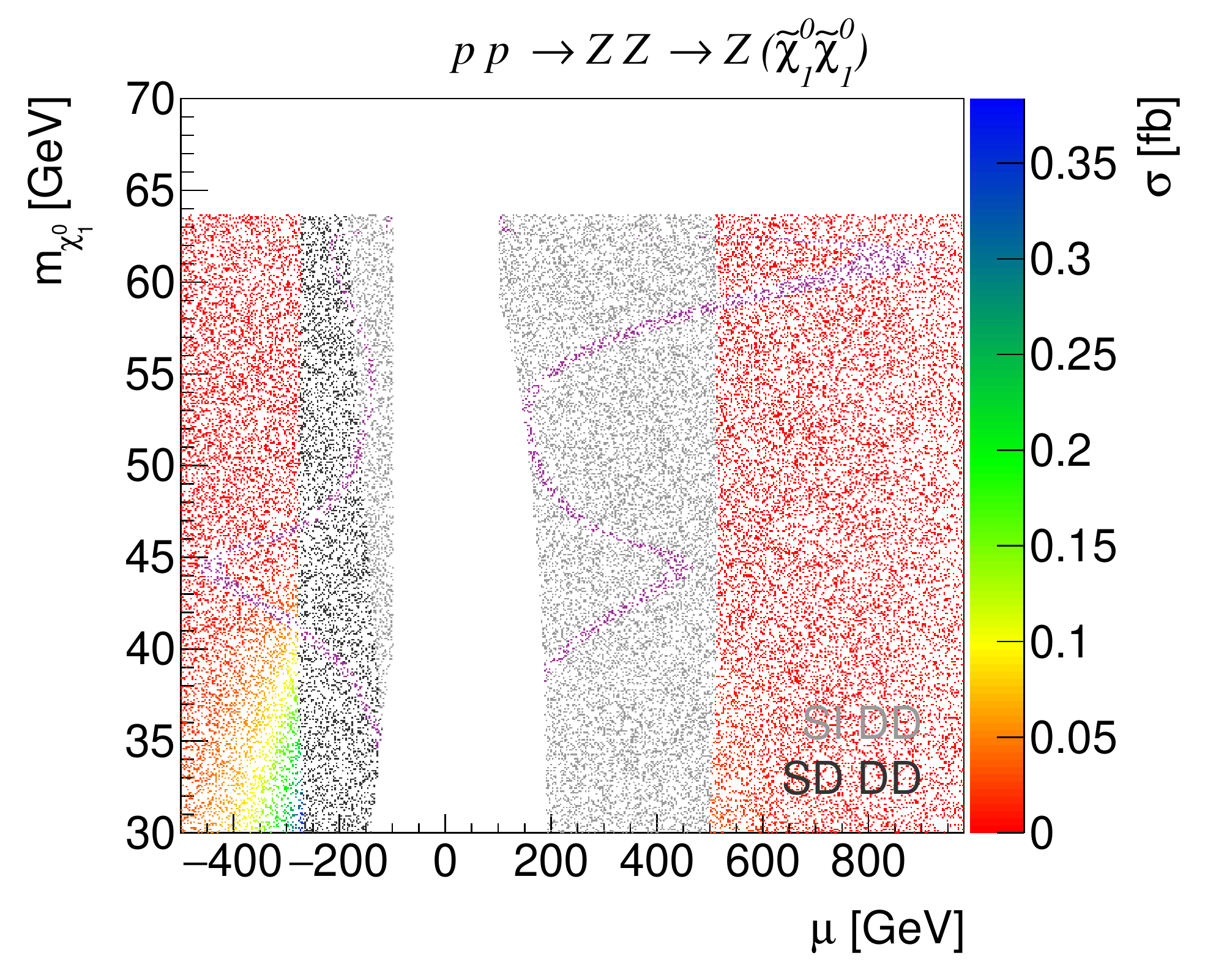} \\
	\includegraphics[width=0.485\textwidth]{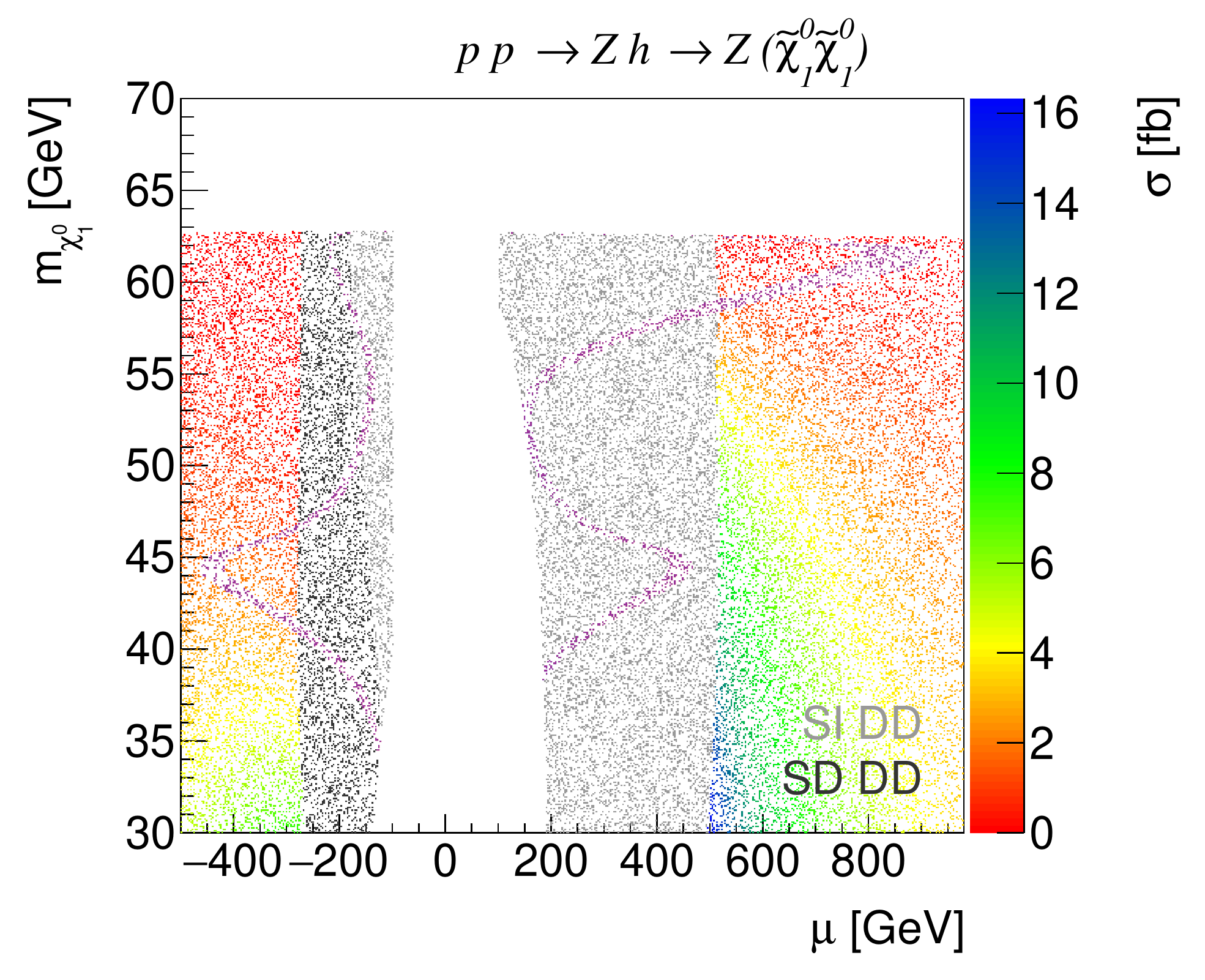} 
	\hspace*{0.02\textwidth}
	\includegraphics[width=0.485\textwidth]{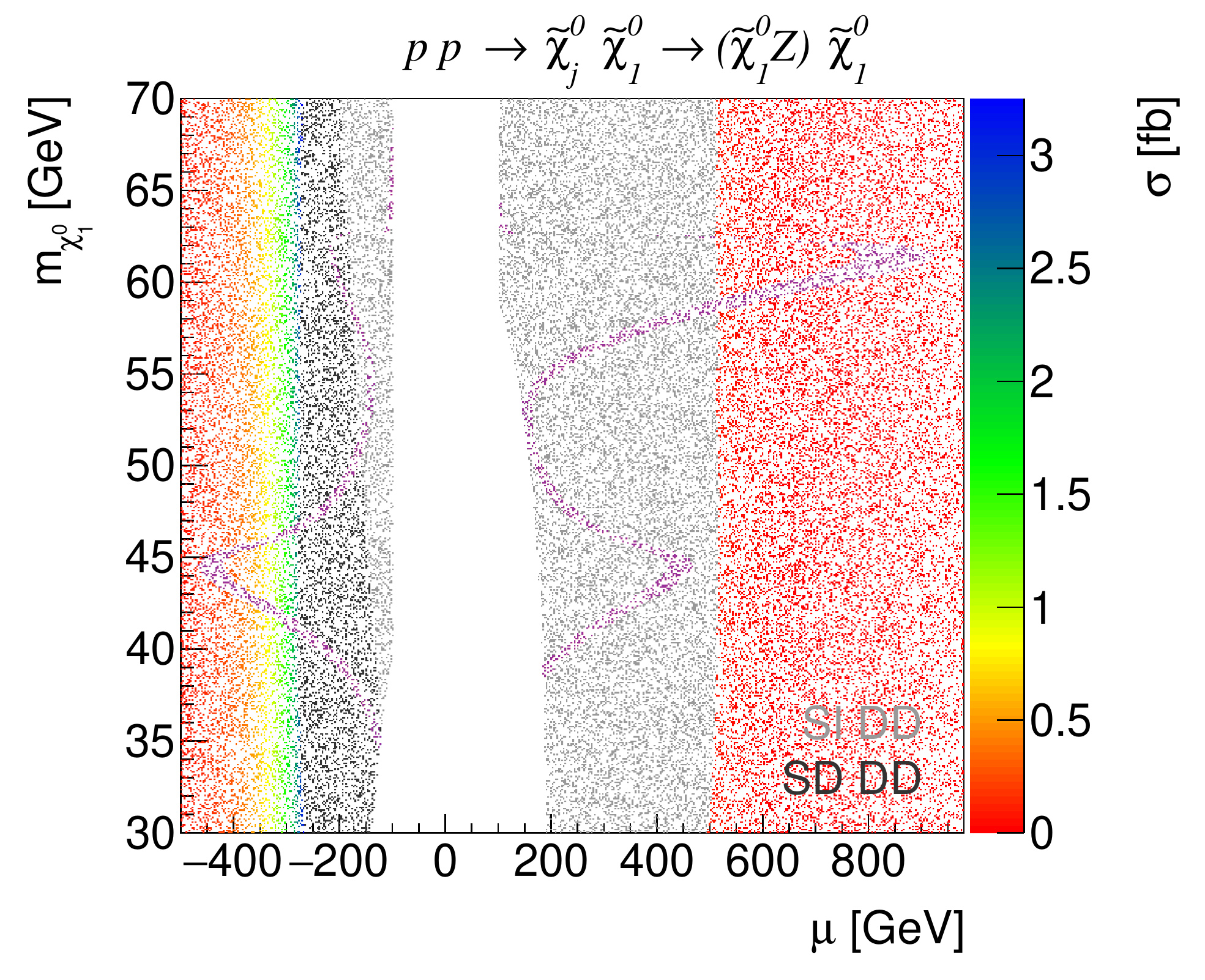}
	\caption{Mono-$Z$ cross sections with DD limits, but without requiring the observed relic density (upper left). Again, we show the three LHC topologies separately (upper right to lower right). Points excluded by spin-independent DD limits are light gray, points excluded by spin-dependent direct detection in dark gray. Points that match the observed relic density under standard assumptions are shown in purple.}
	\label{fig:monozcs_dd_relicindicated}
\end{figure}
%------------------------------------------------------------

As a caveat to the discussion above, we note that including the relic density as a constraint relies on assuming a standard cosmological history. In particular, this requires standard thermal DM production and freeze-out. Further, it assumes that the LSP is the only component of dark matter. Since it is not entirely clear that these assumptions hold, we show mono-$Z$ rates including DD limits, but without imposing the relic density constraint, in Fig.~\ref{fig:monozcs_dd_relicindicated}. Neglecting the relic density allows for significantly larger mono-$Z$ rates. In particular, light DM with mass $\mne{1} \lesssim 40$~GeV strongly enhances the decays $Z\to \nz{1}\nz{1}$ and $h\to\nz{1}\nz{1}$. As a consequence, the $Zh$ topology dominates the mono-$Z$ cross section even more strongly than with the observed relic density and the standard cosmological history imposed.

\bigskip

While we are not arguing that ATLAS and CMS should not perform
mono-$Z$ searches, we have seen that any interpretation of such a
signal as dark matter is likely to require a modification of the
standard thermal freeze-out cosmology. In large parts of the allowed parameter space, the dominant mono-$Z$ topology
in the MSSM, after taking into account all constraints, is invisible
Higgs decays. Those are best searched for in weak-boson-fusion
production~\cite{dieter,anke_jennie}, while mono-$Z$ production can
only confirm the invisible Higgs measurement and add at most very
little new information. So there goes the glory of mono-$Z$.

%%%%%%%%%%%%%%%%%%%%%%%%%%%%%%%%%%%%%%%%%%%%%%%%%%%%%%%%%%%%
\subsection{Mono-W(-pairs)}
\label{sec:fsr_w}

%------------------------------------------------------------
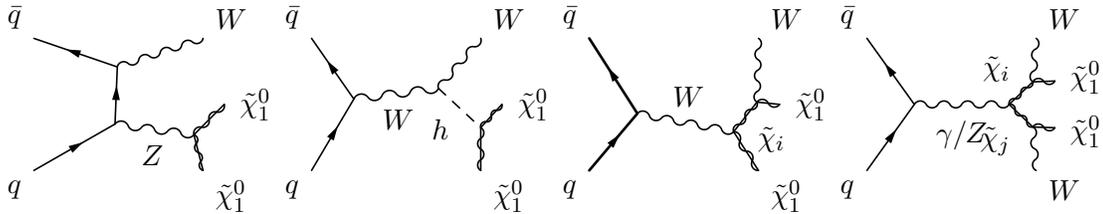
\begin{figure}[b!]
\begin{center}
\begin{fmfgraph*}(80,50)
\fmfset{arrow_len}{2mm}
\fmfleft{i1,i2}
\fmflabel{$q$}{i1}
\fmflabel{$\bar{q}$}{i2}
\fmfright{o1,o2,o3}
\fmflabel{$\tilde{\chi}^0_1$}{o1}
\fmflabel{$\tilde{\chi}^0_1$}{o2}
\fmflabel{$W$}{o3}
\fmf{fermion,width=0.6}{i1,v1,v3,i2}
\fmf{boson,width=0.6}{v3,o3}
\fmf{boson,label=$Z$,width=0.6}{v1,v2}
\fmf{boson,width=0.6}{o1,v2,o2}
\fmf{plain,width=0.6}{o1,v2,o2}
\end{fmfgraph*}
\hspace*{0.04\textwidth}
\begin{fmfgraph*}(80,50)
\fmfset{arrow_len}{2mm}
\fmfleft{i1,i2}
\fmflabel{$q$}{i1}
\fmflabel{$\bar{q}$}{i2}
\fmfright{o1,o2,o3}
\fmflabel{$\tilde{\chi}^0_1$}{o1}
\fmflabel{$\tilde{\chi}^0_1$}{o2}
\fmflabel{$W$}{o3}
\fmf{fermion,width=0.6}{i1,v1,i2}
\fmf{boson,label=$W$,width=0.6}{v1,v2}
\fmf{boson,width=0.6}{v2,o3}
\fmf{dashes,label=$h$,width=0.6}{v2,v3}
\fmf{boson,width=0.6}{o1,v3,o2}
\fmf{plain,width=0.6}{o1,v3,o2}
\end{fmfgraph*}
\hspace*{0.04\textwidth}
\begin{fmfgraph*}(80,50)
\fmfset{arrow_len}{2mm}
\fmfleft{i1,i2}
\fmflabel{$q$}{i1}
\fmflabel{$\bar{q}$}{i2}
\fmfright{o1,o2,o3}
\fmflabel{$\tilde{\chi}^0_1$}{o1}
\fmflabel{$\tilde{\chi}^0_1$}{o2}
\fmflabel{$W$}{o3}
\fmf{fermion}{i1,v1,i2}
\fmf{boson,label=$W$,width=0.6}{v1,v2}
\fmf{boson,width=0.6}{v2,o1}
\fmf{plain,width=0.6}{v2,o1}
\fmf{boson,width=0.6}{v2,v3}
\fmf{plain,label=$\tilde{\chi}_i$,width=0.6}{v2,v3}
\fmf{boson,width=0.6}{v3,o2}
\fmf{plain,width=0.6}{v3,o2}
\fmf{boson,width=0.6}{v3,o3}
\end{fmfgraph*}
\hspace*{0.04\textwidth}
\begin{fmfgraph*}(80,50)
\fmfset{arrow_len}{2mm}
\fmfleft{i1,i2}
\fmflabel{$q$}{i1}
\fmflabel{$\bar{q}$}{i2}
\fmfright{o1,o2,o3,o4}
\fmflabel{$W$}{o1}
\fmflabel{$\tilde{\chi}^0_1$}{o2}
\fmflabel{$\tilde{\chi}^0_1$}{o3}
\fmflabel{$W$}{o4}
\fmf{fermion,width=0.6}{i1,v1,i2}
\fmf{boson,label=$\gamma/Z$,width=0.6}{v1,v2}
\fmf{boson,width=0.6}{v2,v3}
\fmf{plain,label=$\tilde{\chi}_j$,width=0.6}{v2,v3}
\fmf{boson,width=0.6}{v2,v4}
\fmf{plain,label=$\tilde{\chi}_i$,label.side=left,width=0.6}{v2,v4}
\fmf{boson,width=0.6}{v3,o2}
\fmf{plain,width=0.6}{v3,o2}
\fmf{boson,width=0.6}{v4,o3}
\fmf{plain,width=0.6}{v4,o3}
\fmf{boson,width=0.6}{v3,o1}
\fmf{boson,width=0.6}{v4,o4}
\end{fmfgraph*}
\end{center}
\caption{Feynman diagrams contributing to mono-$W$ production in the
  MSSM, including initial-state $W$-radiation with a $Z$-portal, $Wh$
  production with a SM-like Higgs portal, chargino decays, and
  $W$-pair production.}
\label{fig:feyn_monow_mssm}
\end{figure}
%------------------------------------------------------------

Mono-$W$ production is defined through the hard process
\begin{align}
pp \to \nz{1} \nz{1} \; W^\pm \; .
\label{eq:proc_monow_mssm1} 
\end{align}
The relevant MSSM diagrams contributing to this process are shown as
the first three diagrams in Fig.~\ref{fig:feyn_monow_mssm}. Like for
mono-$Z$ production, we can distinguish three topologies,
\begin{alignat}{7}
pp &\to W^\pm Z \to W^\pm  \; (\nz{1} \nz{1}) 
& \qqqquad &\text{ISR} \notag \\
pp &\to W^\pm h \to W^\pm \; (\nz{1} \nz{1}) 
& \qqqquad &\text{invisible Higgs decays} \notag \\
pp &\to \cpm{j} \nz{1} \to (\nz{1} W^\pm) \; \nz{1} 
& \qqqquad &\text{heavy charginos $j=1,2$} \; .
\end{alignat}
The first two rely on the same dark matter couplings as their mono-$Z$
counterparts and only differ in the production process of the SM-like
mediators. Therefore, we will again focus on bino-higgsino dark matter
for the ISR and invisible Higgs topologies.

Also in analogy to mono-$Z$ production, a third topology features
heavy states from the dark matter sector decaying into dark matter and
a weak boson. The heavy state is one of the two charginos with the 
decay $\cpm{j} \to W \nz{1}$. Again, production and decay are mediated
by the same coupling,
\begin{align}
\sigma_{\nz{1} \nz{1} W} \propto \frac{g_{W \nz{1} \cpm{j}}^4}{\Gamma_{\cpm{j}}} \; .
\end{align}
The coupling $g_{W \nz{1} \cpm{j}}$ is in part a higgsino-higgsino
coupling, which following Sec.~\ref{sec:fsr_z} leads us to consider
bino-higgsino dark matter. In addition, $g_{W \nz{1} \cpm{j}}$ includes a
wino-wino interaction. However, bino-wino dark matter is difficult to
reconcile with LEP bounds in the absence of explicit bino-wino mixing
in the neutralino mass matrix. Therefore, all three mono-$W$
topologies again lead us to focus on bino-higgsino dark matter with
\begin{align}
M_1 < \lvert \mu \rvert \ll M_2 \; ,
\end{align}
just like for the mono-$Z$ analysis in Sec.~\ref{sec:fsr_z}.\bigskip

%------------------------------------------------------------
\begin{figure}[t]
\includegraphics[width=0.485\textwidth]{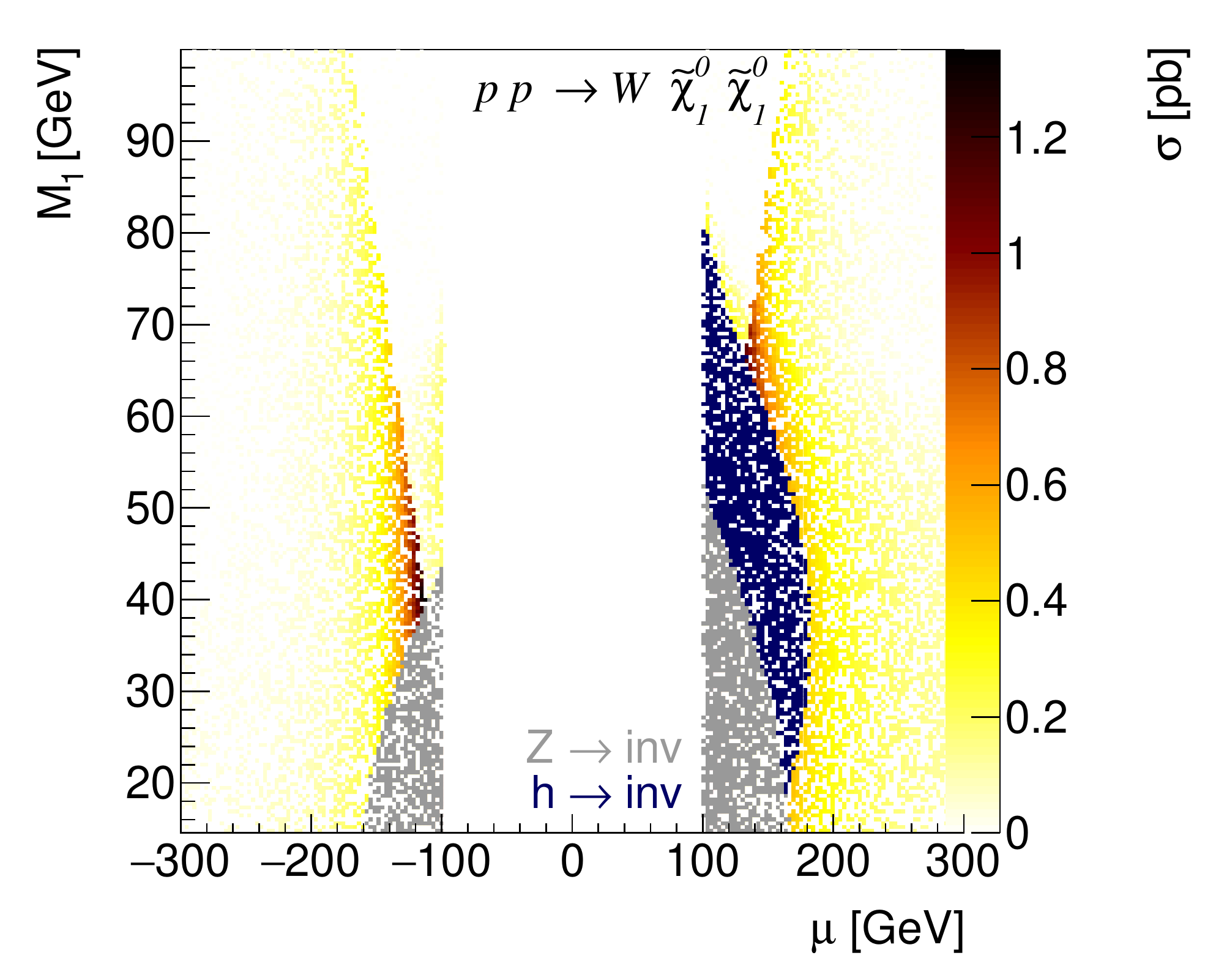}
\hspace*{0.02\textwidth}
\includegraphics[width=0.485\textwidth]{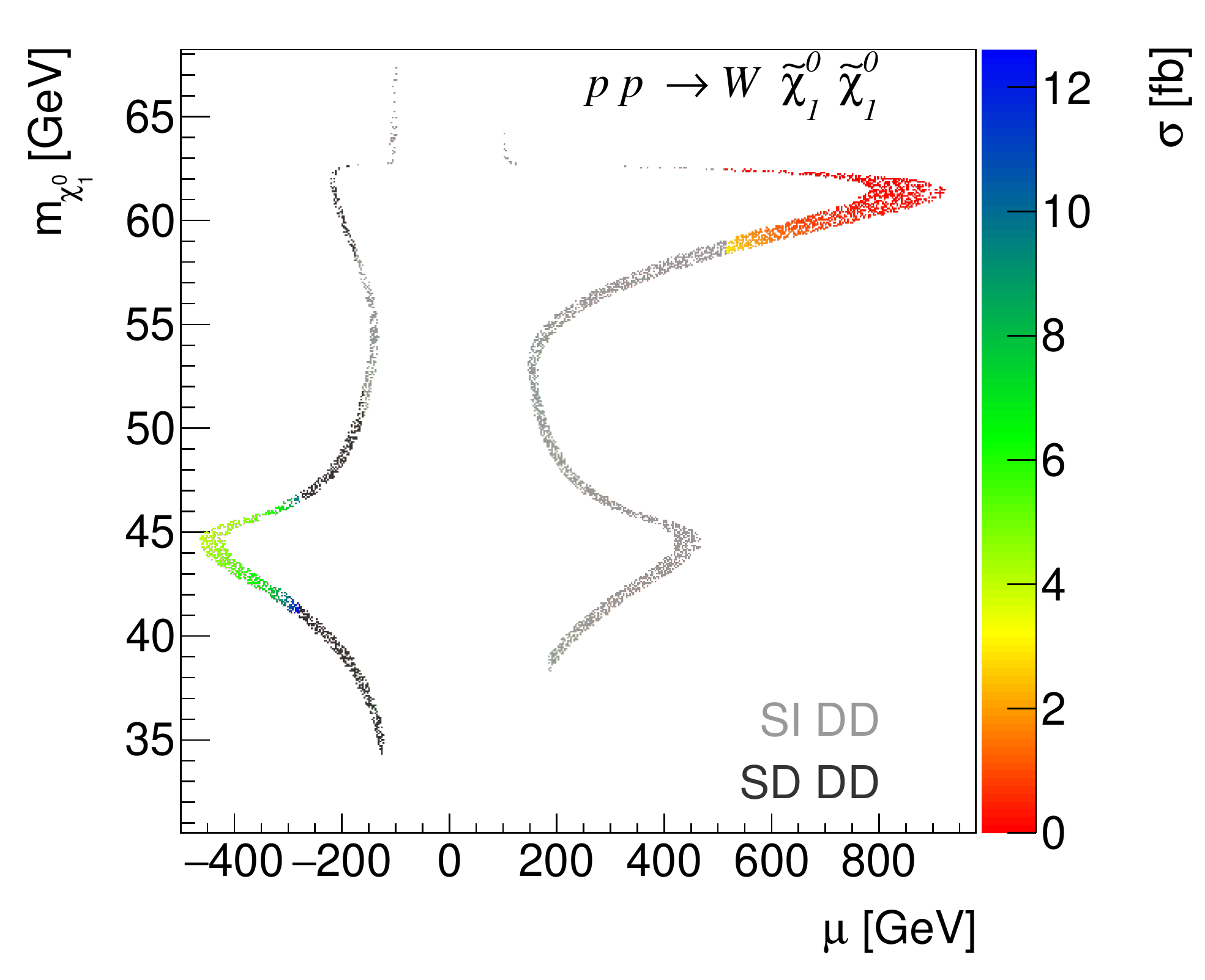}\\
\caption{Cross section for the mono-$W$ process in the $\mu-M_1$
  plane. Left: points fulfilling the chargino mass bound, shown with
  the limits on invisible $Z$ and Higgs decays. Right: points also
  predicting the correct relic density, shown with DD bounds.}
\label{fig:monowcs}
\end{figure}
%------------------------------------------------------------

%------------------------------------------------------------
\begin{figure}[t]
\includegraphics[width=0.485\textwidth]{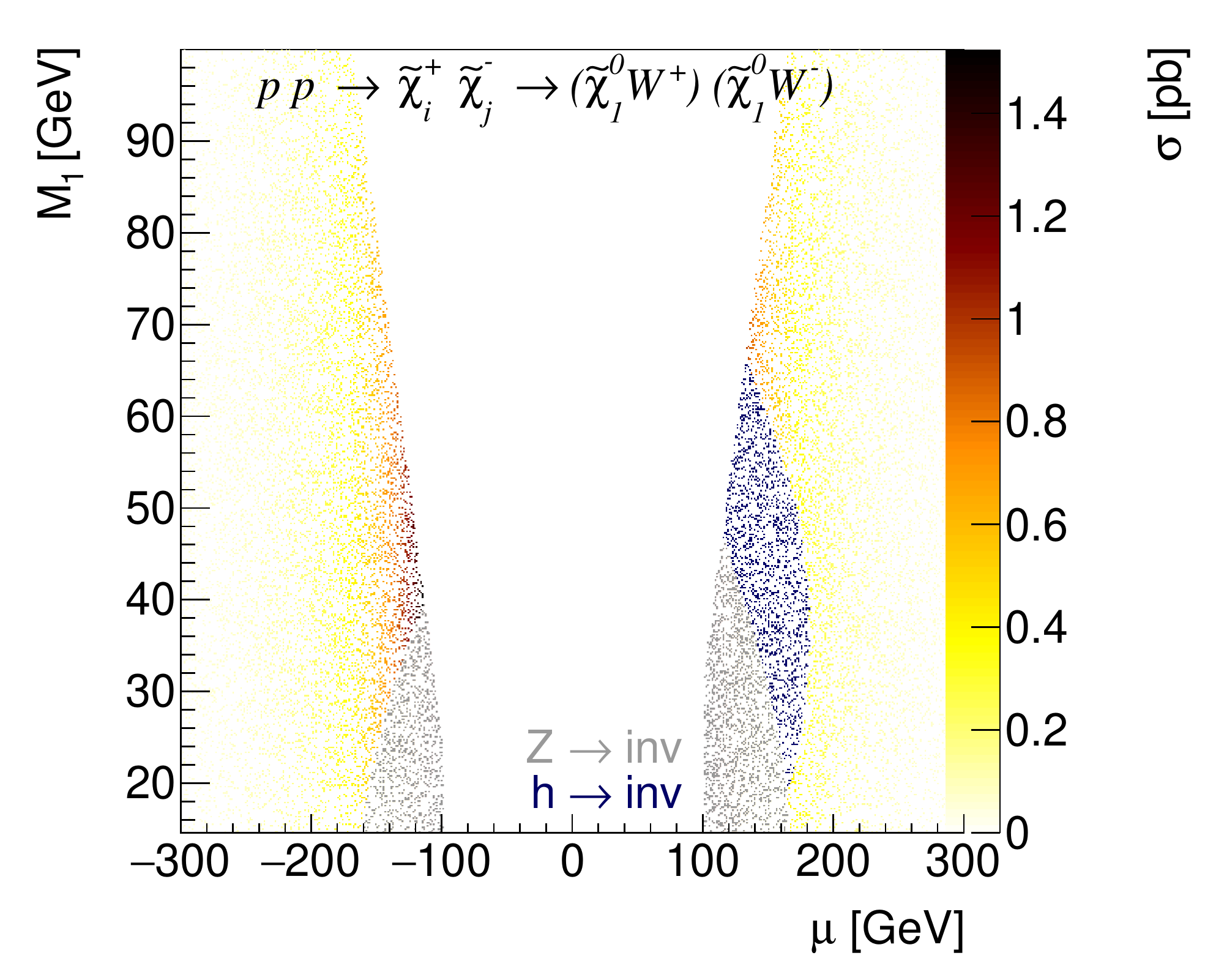}
\hspace*{0.02\textwidth}
\includegraphics[width=0.485\textwidth]{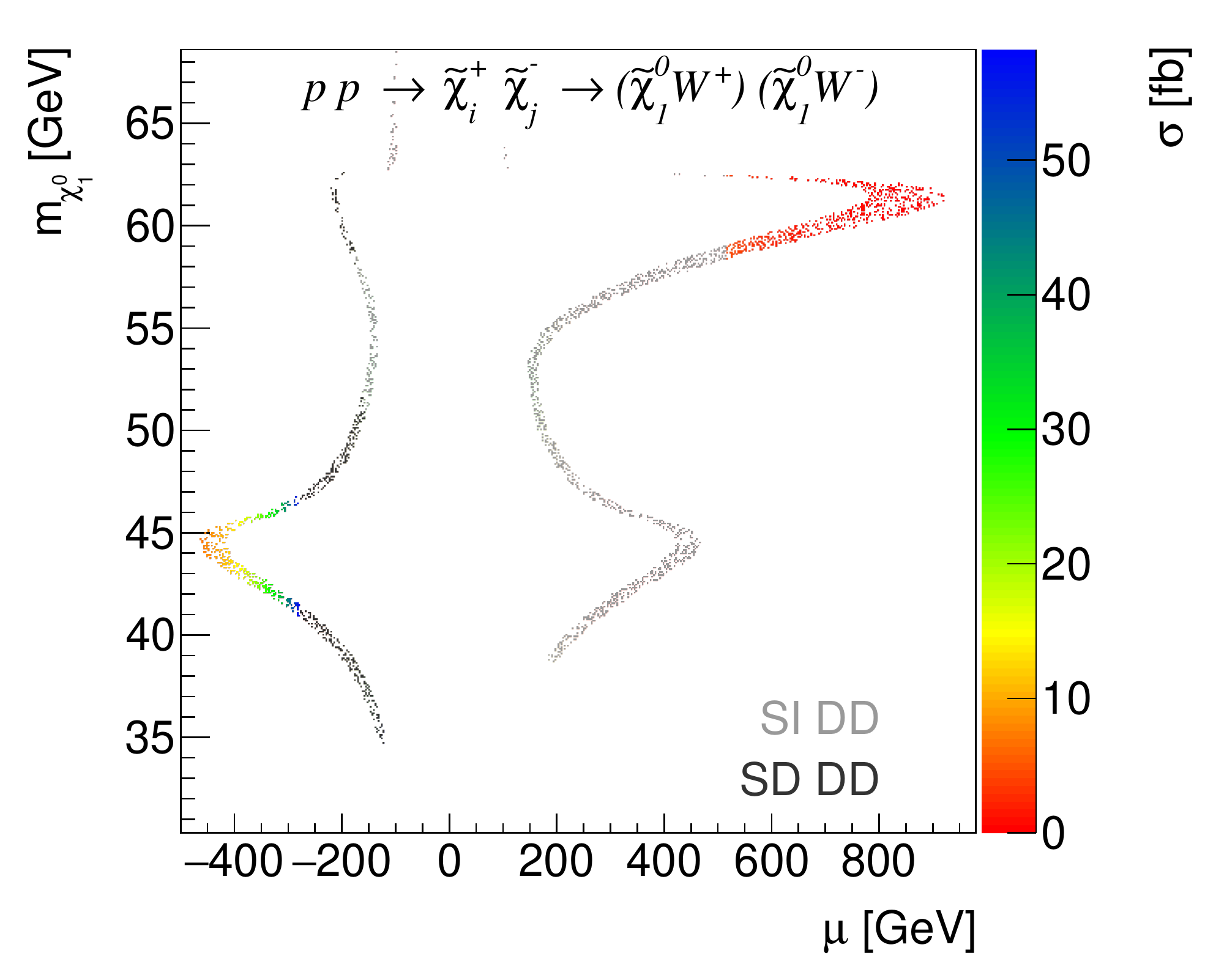}
\caption{Cross section for the mono-$W$-pair process in the
  $\mu-M_1$ plane in analogy to Fig.~\ref{fig:monowcs}. Left: points
  fulfilling the chargino mass bound, shown with the limits on
  invisible $Z$ and Higgs decays. Right: points also predicting the
  correct relic density, shown with DD bounds.}
\label{fig:monodiwcs}
\end{figure}
%------------------------------------------------------------

In the left panel of Fig.~\ref{fig:monowcs} we show the mono-$W$ rate
in the $\mu-M_1$ plane. Like in 
Fig.~\ref{fig:monozcs_lephinv} we again include the limits on invisible
decays. The largest rates lie in the pb range and stem from the
chargino-decay topology. They are two to three times as large as the
largest mono-$Z$ rates passing the same constraints. This is due
partly to the combination of mono-$W^+$ and mono-$W^-$ production,
and partly to the relevant $Z$ and $W$ couplings.

In the right panel of Fig.~\ref{fig:monowcs} we show the points in
agreement with the observed relic density. In addition, we indicate
spin-independent and spin-dependent DD limits. Since the mono-$W$
topologies rely on the same type of dark matter couplings as mono-$Z$
production, the constraints work the same way as in
Sec.~\ref{sec:fsr_z}: ISR rates become negligible, while rates from
chargino decays are suppressed by the large (charged) higgsino masses
required by direct detection.
The largest LHC rates are again found in a narrow window around the
$Z$-pole annihilation funnel. The only difference is that typically
mono-$W$ rates are roughly twice as large as mono-$Z$ rates.\bigskip

A major constraint on mono-$Z$ and mono-$W$ rates at the LHC are DD
limits. Both, spin-independent and spin-dependent DD limits impose a
strong upper bound on the higgsino admixture in the dark matter
candidate through the $g_{Z\nz{1}\nz{1}}$ and $g_{h\nz{1}\nz{1}}$ couplings.
We can try to circumvent them through an LHC production process which
survives the limit $N_{13}, N_{14} \to 0$.  This happens for
mono-$W$-pair production
\begin{align}
pp &\to \cp{i} \cm{j} \to (\nz{1} W^+) \; (\nz{1} W^-) 
\qquad \text{with} \; i,j=1,2 \; ,
\end{align}
shown in the right diagram of Fig.~\ref{fig:feyn_monow_mssm}. The rate
for chargino pair production through an $s$-channel photon is strongly
enhanced compared to purely weak mono-$W$ production. It
does not have a counterpart in mono-$Z$ production. Furthermore, even
for small couplings we can assume
\begin{align}
\br \left(\cpm{1} \to W^\pm \nz{1}\right) \approx 1 \; ,
\end{align}
since it is the only kinematically allowed two-particle decay mode at
tree level.  We show the rates for mono-$W$-pair production in
Fig.~\ref{fig:monodiwcs}. Before taking into account DD constraints,
the rates for mono-$W$ and mono-$W$-pair production are similar. Since
the actual couplings are not constrained by direct detection, the
maximum rates remain larger than for mono-$W$ production. However, we
find that the spin-dependent DD bound on the neutral higgsino, $| \mu
|\gtrsim 250$~GeV leads to a kinematic suppression of the $\cm{j} \cp{j}$
production rate.\bigskip

Our mono-$W$ study implies that in contrast to, for instance,
effective theory arguments, intermediate on-shell states prefer
mono-$W$ production over mono-$Z$ production. One of the mechanisms
behind this is the mono-$W$-pair topology. Its contributions are
removed, if we employ jet or lepton vetoes to remove top backgrounds
for the mono-$W$ signal. Again, there is no point in performing a
detailed signal-background analysis of this channel, because chargino
pair production is a bread-and-butter signature for electroweakinos at
the LHC~\cite{chargino_pairs}.

%%%%%%%%%%%%%%%%%%%%%%%%%%%%%%%%%%%%%%%%%%%%%%%%%%%%%%%%%%%%
\subsection{Mono-Higgs(-pairs)}
\label{sec:fsr_h}

%------------------------------------------------------------
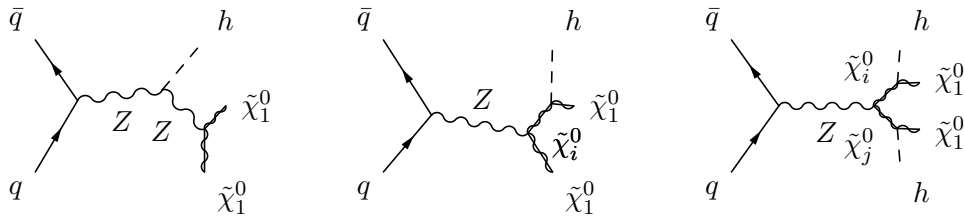
\begin{figure}[b!]
\begin{center}
\begin{fmfgraph*}(80,50)
\fmfset{arrow_len}{2mm}
\fmfleft{i1,i2}
\fmflabel{$q$}{i1}
\fmflabel{$\bar{q}$}{i2}
\fmfright{o1,o2,o3}
\fmflabel{$\tilde{\chi}^0_1$}{o1}
\fmflabel{$\tilde{\chi}^0_1$}{o2}
\fmflabel{$h$}{o3}
\fmf{fermion,width=0.6}{i1,v1,i2}
\fmf{boson,label=$Z$,width=0.6}{v1,v2}
\fmf{dashes,width=0.6}{v2,o3}
\fmf{boson,label=$Z$,width=0.6}{v2,v3}
\fmf{boson,width=0.6}{o1,v3,o2}
\fmf{plain,width=0.6}{o1,v3,o2}
\end{fmfgraph*}
\hspace*{0.1\textwidth}
\begin{fmfgraph*}(80,50)
\fmfset{arrow_len}{2mm}
\fmfleft{i1,i2}
\fmflabel{$q$}{i1}
\fmflabel{$\bar{q}$}{i2}
\fmfright{o1,o2,o3}
\fmflabel{$\tilde{\chi}^0_1$}{o1}
\fmflabel{$\tilde{\chi}^0_1$}{o2}
\fmflabel{$h$}{o3}
\fmf{fermion,width=0.6}{i1,v1,i2}
\fmf{boson,label=$Z$,width=0.6}{v1,v2}
\fmf{boson,width=0.6}{v2,o1}
\fmf{plain,width=0.6}{v2,o1}
\fmf{boson,label=$\tilde{\chi}^0_i$,width=0.6}{v2,v3}
\fmf{plain,label=$\tilde{\chi}^0_i$,width=0.6}{v2,v3}
\fmf{boson,width=0.6}{v3,o2}
\fmf{plain,width=0.6}{v3,o2}
\fmf{dashes,width=0.6}{v3,o3}
\end{fmfgraph*}
\hspace*{0.1\textwidth}
\begin{fmfgraph*}(80,50)
\fmfset{arrow_len}{2mm}
\fmfleft{i1,i2}
\fmflabel{$q$}{i1}
\fmflabel{$\bar{q}$}{i2}
\fmfright{o1,o2,o3,o4}
\fmflabel{$h$}{o1}
\fmflabel{$\tilde{\chi}^0_1$}{o2}
\fmflabel{$\tilde{\chi}^0_1$}{o3}
\fmflabel{$h$}{o4}
\fmf{fermion,width=0.6}{i1,v1,i2}
\fmf{boson,label=$Z$,width=0.6}{v1,v2}
\fmf{boson,width=0.6}{v2,v3}
\fmf{plain,label=$\tilde{\chi}^0_j$,width=0.6}{v2,v3}
\fmf{boson,width=0.6}{v2,v4}
\fmf{plain,label=$\tilde{\chi}^0_i$,label.side=left,width=0.6}{v2,v4}
\fmf{boson,width=0.6}{v3,o2}
\fmf{plain,width=0.6}{v3,o2}
\fmf{boson,width=0.6}{v4,o3}
\fmf{plain,width=0.6}{v4,o3}
\fmf{dashes,width=0.6}{v3,o1}
\fmf{dashes,width=0.6}{v4,o4}
\end{fmfgraph*}
\end{center}
  \caption{Feynman diagrams contributing to mono-Higgs production in
    the MSSM, $Zh$ production with a $Z$-portal, and heavy neutralino
    decays, and Higgs pair production.}
  \label{fig:feyn_monoh_mssm}
\end{figure}
%------------------------------------------------------------

Mono-Higgs production is the third electroweak process we consider in
our comprehensive study of final state decay leading to mono-$X$
signatures. The hard process reads
\begin{align}
pp \to \nz{1} \nz{1} h \; .
\label{eq:proc_monoh_mssm} 
\end{align}
The Higgs boson in the final state is the SM-like light scalar of the
MSSM.  The Feynman diagrams shown in Fig.~\ref{fig:feyn_monoh_mssm}
define two mono-Higgs topologies
\begin{alignat}{7}
pp &\to hZ \to h \; (\nz{1} \nz{1}) 
& \qqqquad &\text{invisible $Z$-decays} \notag \\
pp &\to \nz{j} \nz{1} \to (\nz{1} h) \; \nz{1} 
& \qqqquad &\text{heavy neutralinos $j=2,3,4$} \; .
\end{alignat}
Obviously, the usual ISR topology is not relevant for the Higgs
case. The $Zh$ topology is based on the same production mechanism as
for mono-$Z$ production, but combined with a strongly constrained
branching ratio $\br_{Z \to \chi \chi}$.  The two relevant couplings
driving the neutralino decay topology are
\begin{align}
\sigma_{\nz{1} \nz{1} h} \propto \frac{g_{Z \nz{1} \nz{i}}^2 g_{h \nz{1} \nz{i}}^2}{\Gamma_{\nz{i}}} \; .
\end{align}
The production process still requires a sizable coupling to the
$Z$, while the decay proceeds through the Higgs coupling. The decay
$\nz{i} \to \nz{1} h$ competes with the decay $\nz{i} \to \nz{1} Z$. Just like
for mono-$Z$ and mono-$W$ production, the observed relic density
combined with all available constraints motivates mixed bino-higgsino
dark matter,
\begin{align}
M_1 < \lvert \mu \rvert \ll M_2 \; .
\end{align}
\bigskip

\begin{figure}[t]
\includegraphics[width=0.485\textwidth]{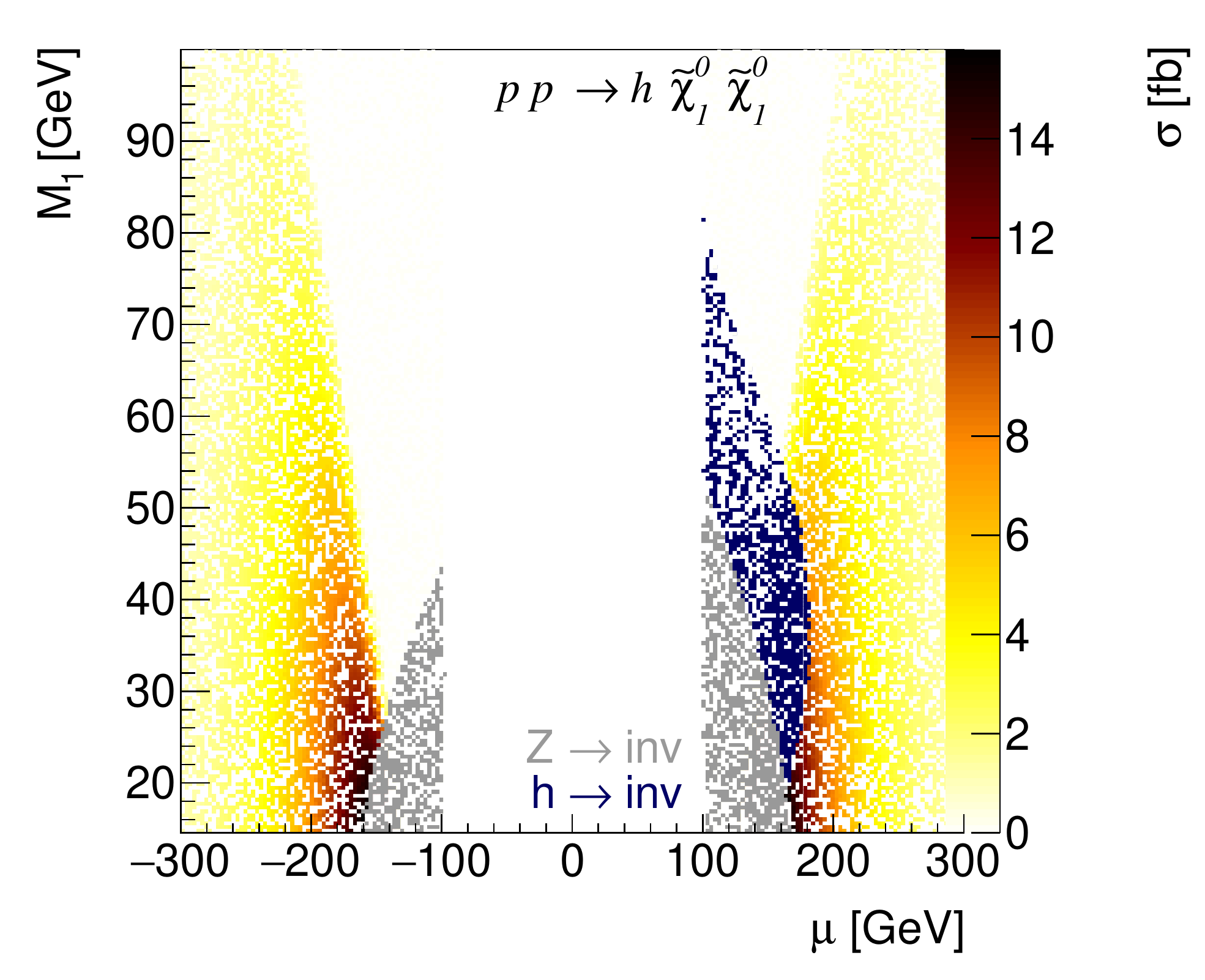}
\hspace*{0.02\textwidth}
\includegraphics[width=0.485\textwidth]{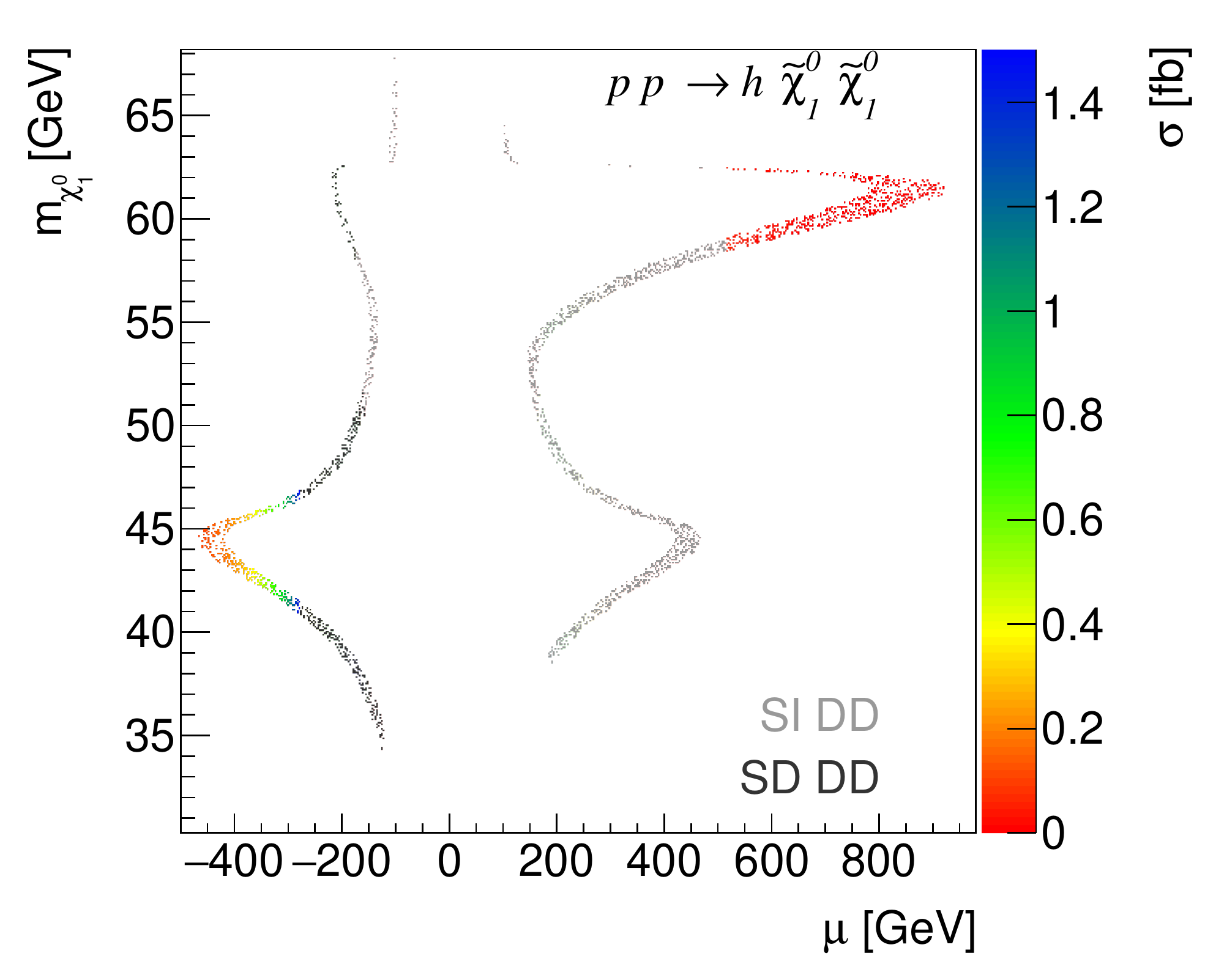}
\caption{Cross section for the mono-Higgs process in the $\mu-M_1$
  plane. Left: points fulfilling the chargino mass bound, shown with
  the limits on invisible $Z$ and Higgs decays. Right: points also
  predicting the correct relic density, shown with DD bounds.}
\label{fig:monoh_cs}
\end{figure}

In the left panel of Fig.~\ref{fig:monoh_cs} we show the rates we
start with, before considering relic density and DD constraints. We
see that the mono-Higgs rates are more than an order of magnitude
smaller than their mono-$Z$ or mono-$W$ counterparts shown in
Fig.~\ref{fig:monozcs_lephinv} and Fig.~\ref{fig:monowcs}. For the
$Zh$ topology the limiting factor is the smaller invisible branching
ratio of the $Z$-boson as compared to the invisible Higgs decays,
described in Sec.~\ref{sec:dm_sfitter}.  The neutralino decay topology
predicts smaller rates because especially for large production rates
and before considering DD limits the competing decay rate $\nz{2,3}
\to \nz{1} Z$ is large.

%------------------------------------------------------------
%------------------------------------------------------------

In the right panel of Fig.~\ref{fig:monoh_cs} we see the effect of the
spin-dependent and spin-independent DD limits. The $Zh$ topology is
now suppressed to unobservable LHC rates through the invisible $Z$
branching ratio, just like the ISR topology of the mono-$Z$ signature
described in Sec.~\ref{sec:fsr_z}. Unlike for the mono-$Z$ case, the
neutralino decay topology becomes the leading channel with possible
LHC rates in the range of 1~fb. The predicted mono-Higgs rate after
taking into account DD constraints is indeed not much smaller than the
expected mono-$Z$ rates from neutralino decay.\bigskip

Inspired by the mono-$W$ case, it turns out that one way out of some
of the leading constraints is mono-Higgs-pair production shown in the
right panel of Fig.~\ref{fig:feyn_monoh_mssm},
\begin{align}
pp \to \nz{i} \nz{j} \to ( \nz{1} h ) \; ( \nz{1} h ) 
\qquad \text{with} \; i,j=2,3,4 \; .
\label{eq:proc_monohh_mssm} 
\end{align}
The neutralino production couplings are now separated from the decay
couplings and, more importantly, from the couplings mediating direct
detection,
\begin{align}
\sigma_{\nz{1} \nz{1} hh} \propto \frac{g_{Z \nz{i} \nz{j}}^2 g_{h \nz{1} \nz{i}}^2 g_{h \nz{1} \nz{j}}^2}{\Gamma_{\nz{i}} \Gamma_{\nz{j}}} \; .
\end{align}
In our preferred scenario with bino-higgsino dark matter and another,
relatively light higgsino the production of heavy neutralino pairs
will be sizable. At the same time, the decay to Higgs bosons requires
a gaugino content just like the annihilation responsible for the
correct relic density. 

\begin{figure}[t]
\includegraphics[width=0.485\textwidth]{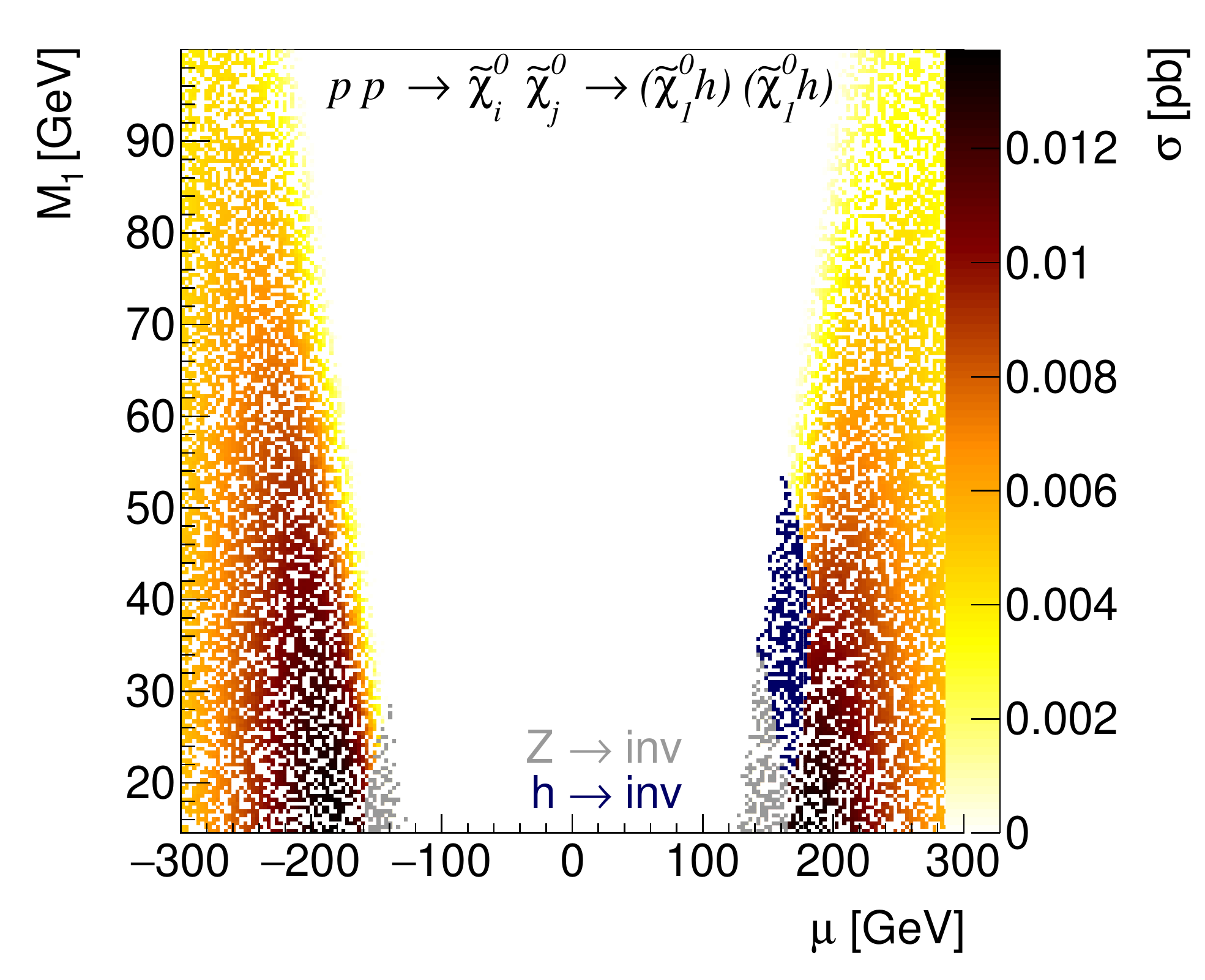}
\hspace*{0.02\textwidth}
\includegraphics[width=0.485\textwidth]{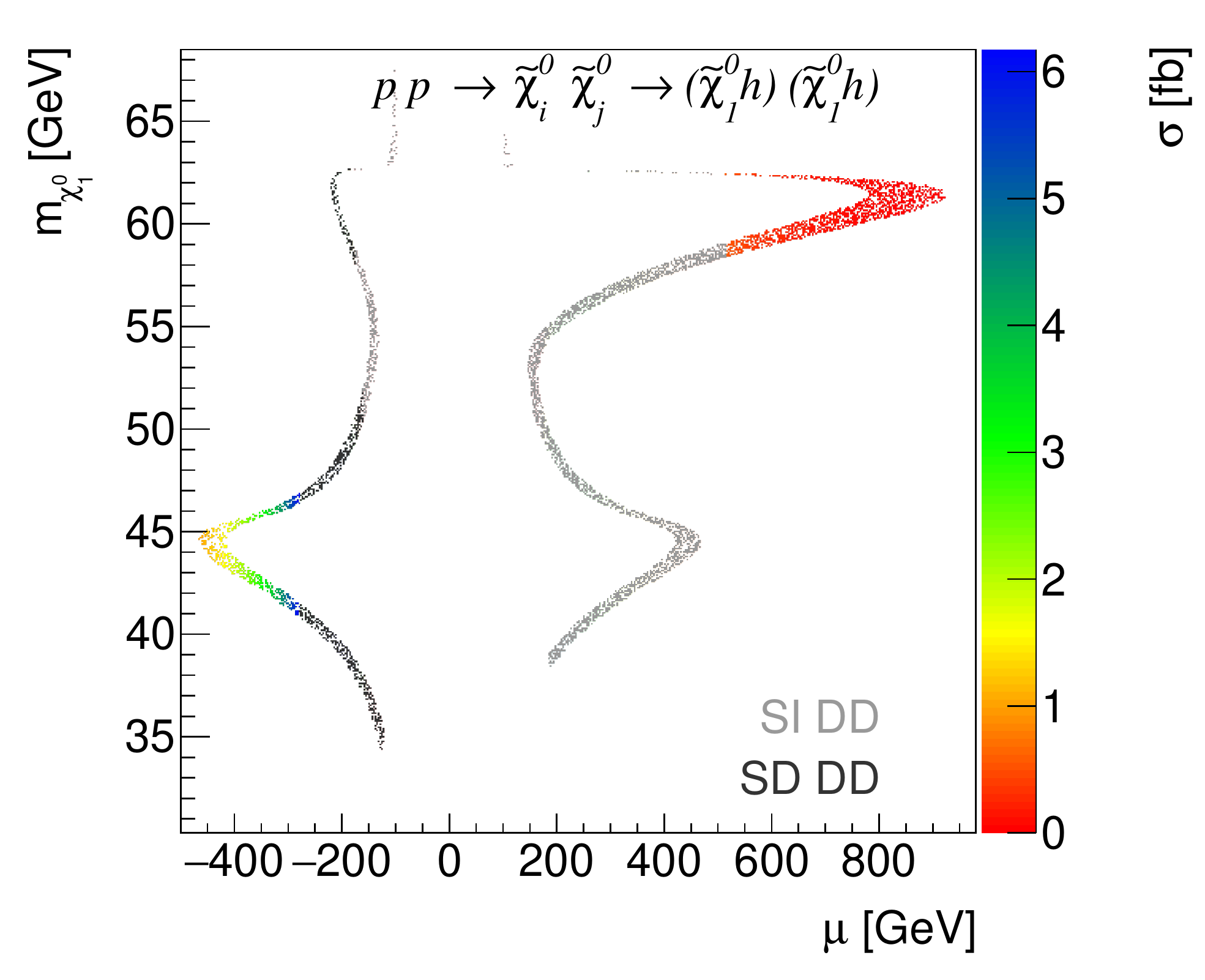}
\caption{Cross section for the mono-Higgs-pair process in the
  $\mu-M_1$ plane. Left: points fulfilling the chargino mass bound,
  shown with the limits on invisible $Z$ and Higgs decays. Right:
  points also predicting the correct relic density, shown with DD
  bounds.}
\label{fig:monodih_cs}
\end{figure}

We show the LHC rates for the mono-Higgs-pair signature in
Fig.~\ref{fig:monodih_cs}.  First, the mono-Higgs-pair cross section
is suppressed by the phase space of two heavy higgsinos in the final
state with $\lvert\mu\rvert\gtrsim 300~...~400$~GeV, just like the
mono-$W$-pair rate. This is why the rate before applying any
constraints is in the same range as the mono-Higgs rate.  On the other
hand, every coupling contributing to the LHC rate is unrelated to
direct detection. Through a large bino fraction of the dark matter
agent we can essentially decouple the DD constraints, so the LHC rates
with and without relic density and DD constraints are very
similar. All we need to do is enhance the annihilation rate in the
early universe through an on-shell condition $\mne{1} \approx M_1
\approx m_Z/2$ or $\mne{1} \approx M_1 \approx m_h/2$.\bigskip

%------------------------------------------------------------
%------------------------------------------------------------

The LHC signature of mono-Higgs-pair production is similar to Higgs
pair production at the LHC.  While the expected production rate for a
pair of SM-like Higgs bosons is around 35~fb, the additional missing
energy in the mono-Higgs-pair signal of Eq.\eqref{eq:proc_monohh_mssm}
should allow for a better background rejection.  Which decay
combination of the two Higgs bosons works best for this purpose is
currently under study~\cite{susanne}.  For SM-like Higgs pairs the
combination $b\bar{b} \; \gamma \gamma$ works best to guarantee
detection and reduce backgrounds, but for the smaller dark matter
signal the combinations $b\bar{b} \; b\bar{b} \; \met$ or $b\bar{b} \;
WW \; \met$ might be more promising.

\begin{figure}[t]
	\includegraphics[width=0.485\textwidth]{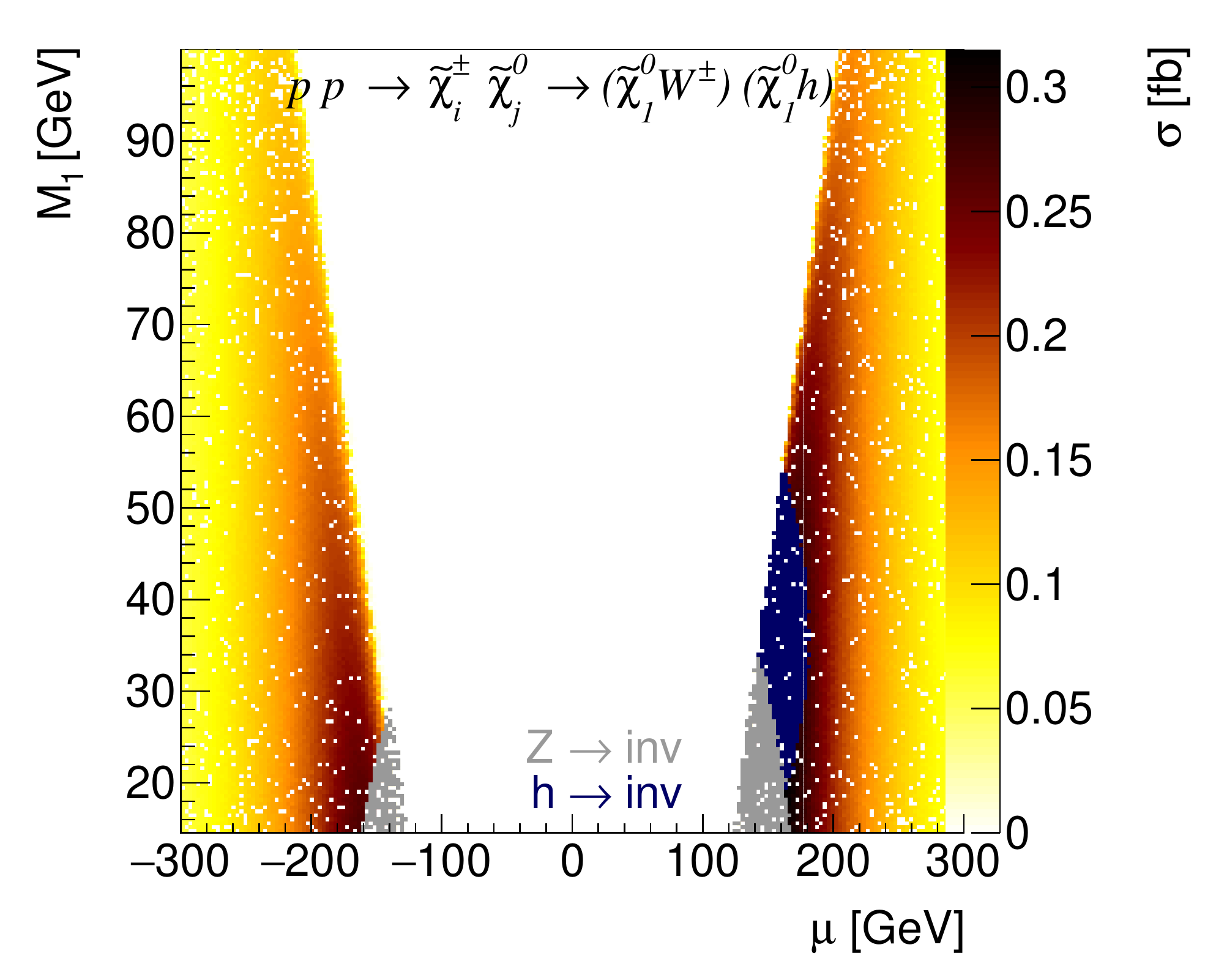}
	\hspace*{0.02\textwidth}
	\includegraphics[width=0.485\textwidth]{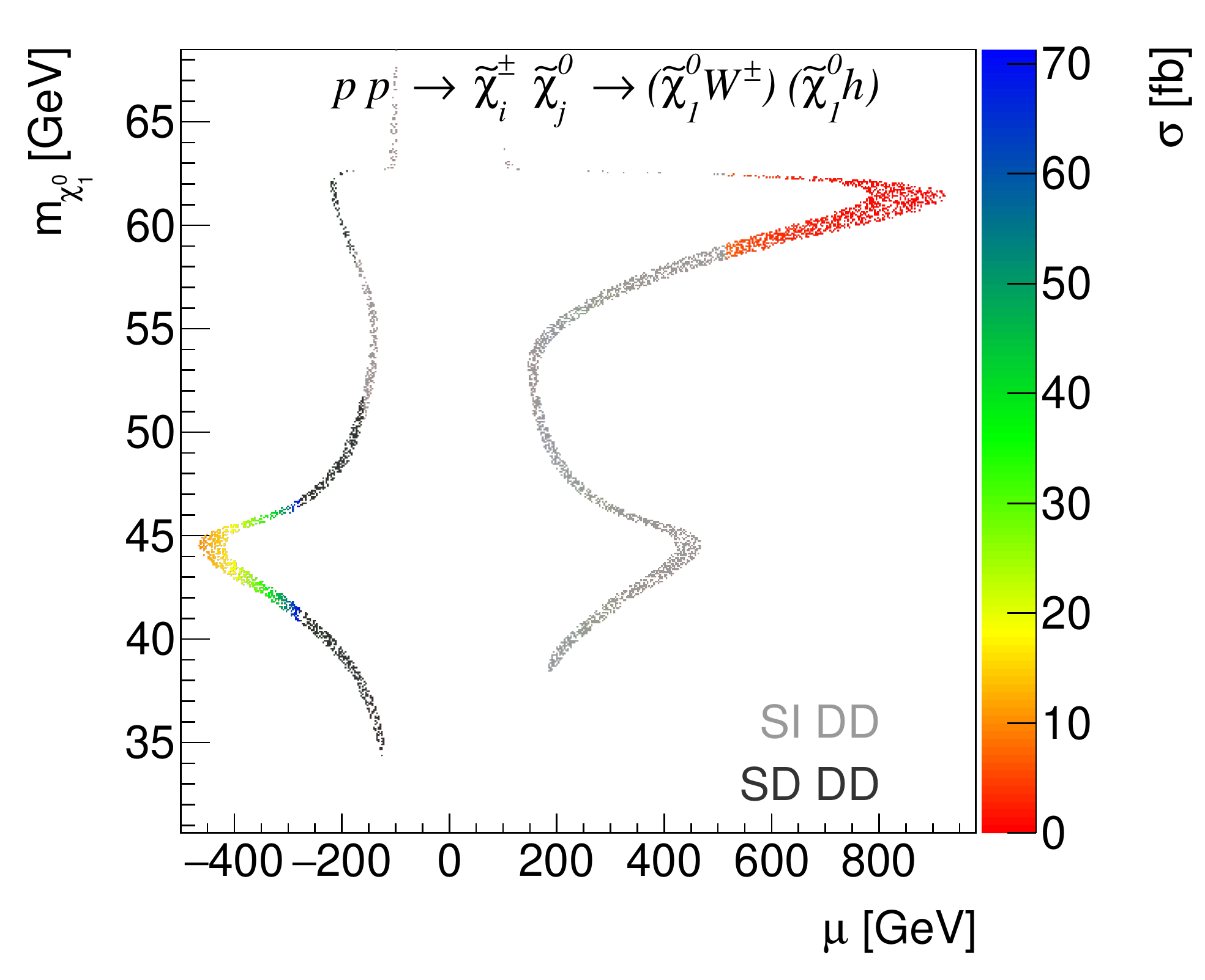}
	\caption{Cross section for the mono-$Wh$-pair process in the
		$\mu-M_1$ plane. Left: points fulfilling the chargino mass bound,
		shown with the limits on invisible $Z$ and Higgs decays. Right:
		points also predicting the correct relic density, shown with DD
		bounds.}
	\label{fig:monowh_cs}
\end{figure}

Finally, for completeness and in analogy to mono-$W$ pairs and mono-Higgs pairs, we consider the mono-$Wh$ pair process given by
\begin{align}
	pp \to \cpm{i} \nz{j} \to ( \nz{1} W^\pm ) \; ( \nz{1} h ) 
	\qquad \text{with} \; i=1,2 \; \text{and} \; j=2,3,4 \; .
	\label{eq:proc_monowh_mssm} 
\end{align}
This topology is driven by the production coupling $g_{W \nz{j} \cpm{i}}$ and, for the decay, $g_{W \nz{1} \cpm{i}}$ and $g_{h \nz{1} \nz{j}}$,
\begin{align}
	\sigma_{\nz{1} \nz{1} Wh} \propto \frac{g_{W \nz{j} \cpm{i}}^2 g_{W \nz{1} \cpm{i}}^2 g_{h \nz{1} \nz{j}}^2}{\Gamma_{\cpm{i}} \Gamma_{\nz{j}}} \; .
\end{align}
In the scenario of a bino-higgsino LSP, heavier higgsinos and a decoupled wino, the production cross section for a heavy chargino-neutralino pair will be sizable. Like in the mono-$W$(-pair) process, we again have
\begin{align}
	\br \left(\cpm{1} \to W^\pm \nz{1}\right) \approx 1 \; .
\end{align}
On the other hand, the decay $\nz{j}\to \nz{1}h$ of the heavy neutralino, requiring gaugino and higgsino parts in the LSP, competes with the decay $\nz{j}\to \nz{1}h$.

Hence, rates before relic density and direct detection constraints, shown in the left panel of Fig.~\ref{fig:monowh_cs}, lie between the rates for mono-Higgs pairs and those for mono-$W$ pairs. Like for mono-$W$ pairs and mono-Higgs pairs, production couplings and decays are decoupled from direct detection. The correct relic density can be guaranteed through the resonant enhancement at $\mne{1} \approx M_1 \approx m_Z/2$ or $\mne{1} \approx M_1 \approx m_h/2$. Hence, the mono-$Wh$-pair cross section is only suppressed kinematically through the production of heavy higgsinos. The resulting rates are shown in the right panel of Fig.~\ref{fig:monowh_cs}. We find cross sections of up to $70$~fb, slightly above the mono-$W$-pair rate.
We do not perform a signal-background analysis, since neutralino-chargino pairs belong to the electroweakino signatures already being studied at the LHC~\cite{chargino_neutralino_pairs}.

%%%%%%%%%%%%%%%%%%%%%%%%%%%%%%%%%%%%%%%%%%%%%%%%%%%%%%%%%%%%
\section{Final state decays beyond the MSSM}
\label{sec:beyond}

The leading constraint on the size of electroweak mono-$X$ signals in
the MSSM comes from direct detection or, more specifically, from the
combination of the relic density constraint and the DD limits. The
reason is that we need large couplings to the $Z$ and SM-like Higgs
mediators to reach the observed relic density, direct detection
strongly constrains these couplings, and most LHC rates again rely on
the same couplings. In extended models like the NMSSM a dark sector
mediator is responsible for the correct relic density, in spite of
very small couplings to the Standard Model. From our mono-$W$(-pair)
and mono-Higgs(-pair) we know how to decouple the decay topologies at
the LHC from the DD constraints, which motivates our NMSSM study.

Following Sec.~\ref{sec:dm_nmssm} we adjust the singlet--singlino dark
matter sector such that a light singlino with $\mne{1}=10$~GeV can
annihilate to the correct relic density through an on-shell singlet.
Because this annihilation relies on the couplings within the
singlet--singlino sector we can decouple the gaugino masses in our
$|\mu| \ll M_1 = M_2 = 1$~TeV.  Following Eq.\eqref{eq:singlet_mass}
and Eq.\eqref{eq:decoup_higgs} we ensure the corresponding mass
relation by choosing $\tilde \kappa$ such that
\begin{align}
\mne{1} \approx 
2\tilde{\kappa}\mu + \frac{m_Z^2}{\mu} \;  \tilde \lambda^2 \; \frac{2\tilde{\kappa}-s_{2\beta}}{4\tilde{\kappa}^2-1}
= 10~\gev \; .
\label{eq:nmssm_lspmass}
\end{align}
If we include the LEP constraints $\lvert\mu\rvert\gtrsim 100$~GeV,
this typically implies
\begin{align}
\lvert\tilde{\kappa}\rvert = \frac{m_{\nz{1}}}{2\lvert\mu\rvert} \lesssim 0.05
\; .
\end{align}
or $\lvert\kappa\rvert \ll \lvert\lambda\rvert$ in the original
notation. For our mass hierarchy this means
\begin{align}
|\tilde{\kappa} \mu| \ll |\mu| \ll M_1 \approx M_2 \; .
\end{align}
The singlino couplings from Eq.\eqref{eq:nmssm_coupA_simp} are
approximately given by
\begin{align}
g_{a_s \nz{1} \nz{1}}
\approx g_{h_s \nz{1} \nz{1}}
\approx - \sqrt{2}  g \; \tilde \lambda \tilde \kappa  \; N_{15}^2  \; .
\label{eq:nmssm_coupA_simp2}
\end{align}
They are not large compared for example to gauge couplings, but
sufficiently large to explain the observed relic density for an
on-shell annihilation process.  The remaining free parameters in the
NMSSM electroweakino sector which we vary in our analysis are $\tilde
\kappa$, $\tilde \lambda$ and $A_\kappa$.\bigskip

%------------------------------------------------------------
\begin{figure}[t]
\includegraphics[width=0.485\textwidth]{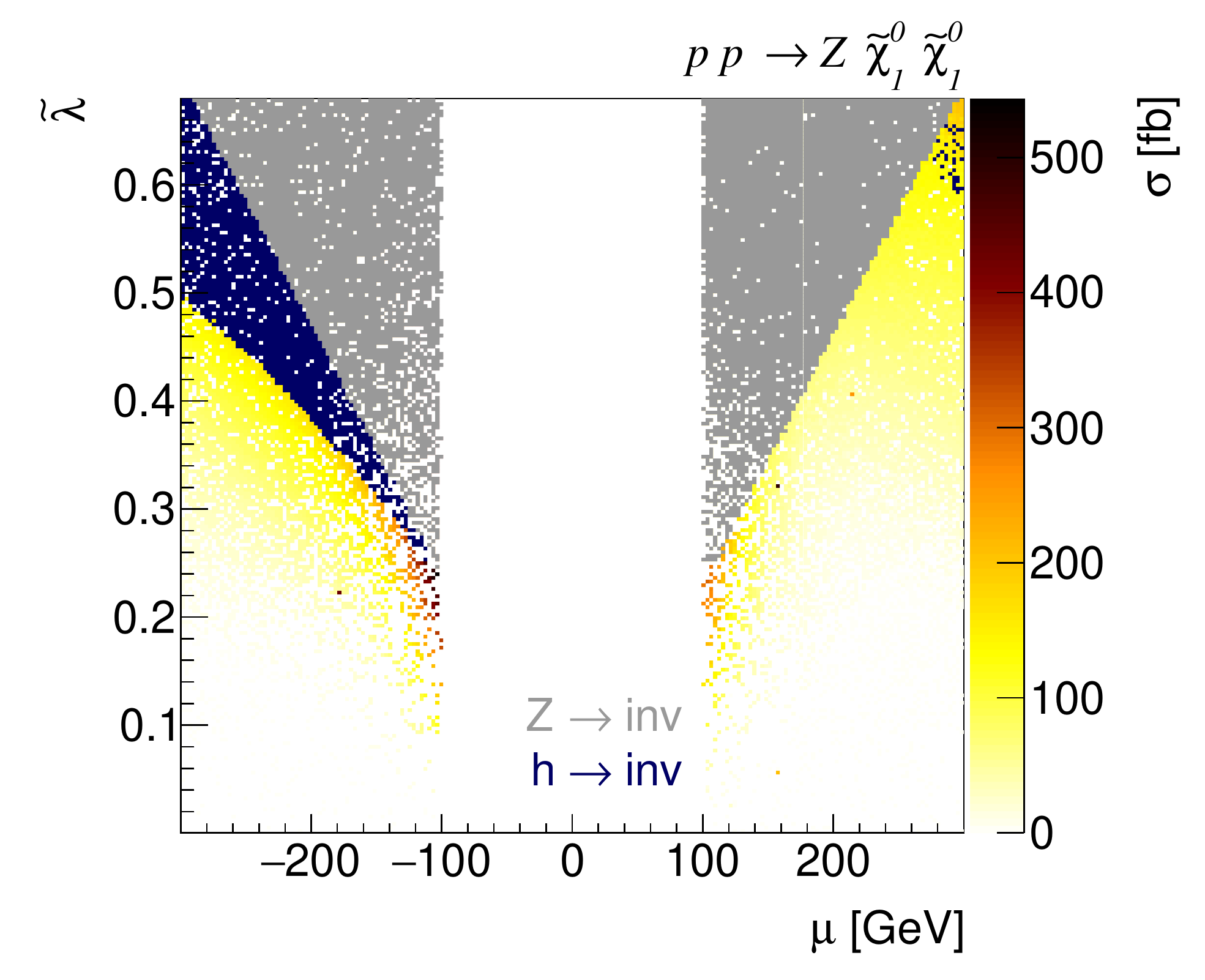}
\hspace*{0.02\textwidth}
\includegraphics[width=0.485\textwidth]{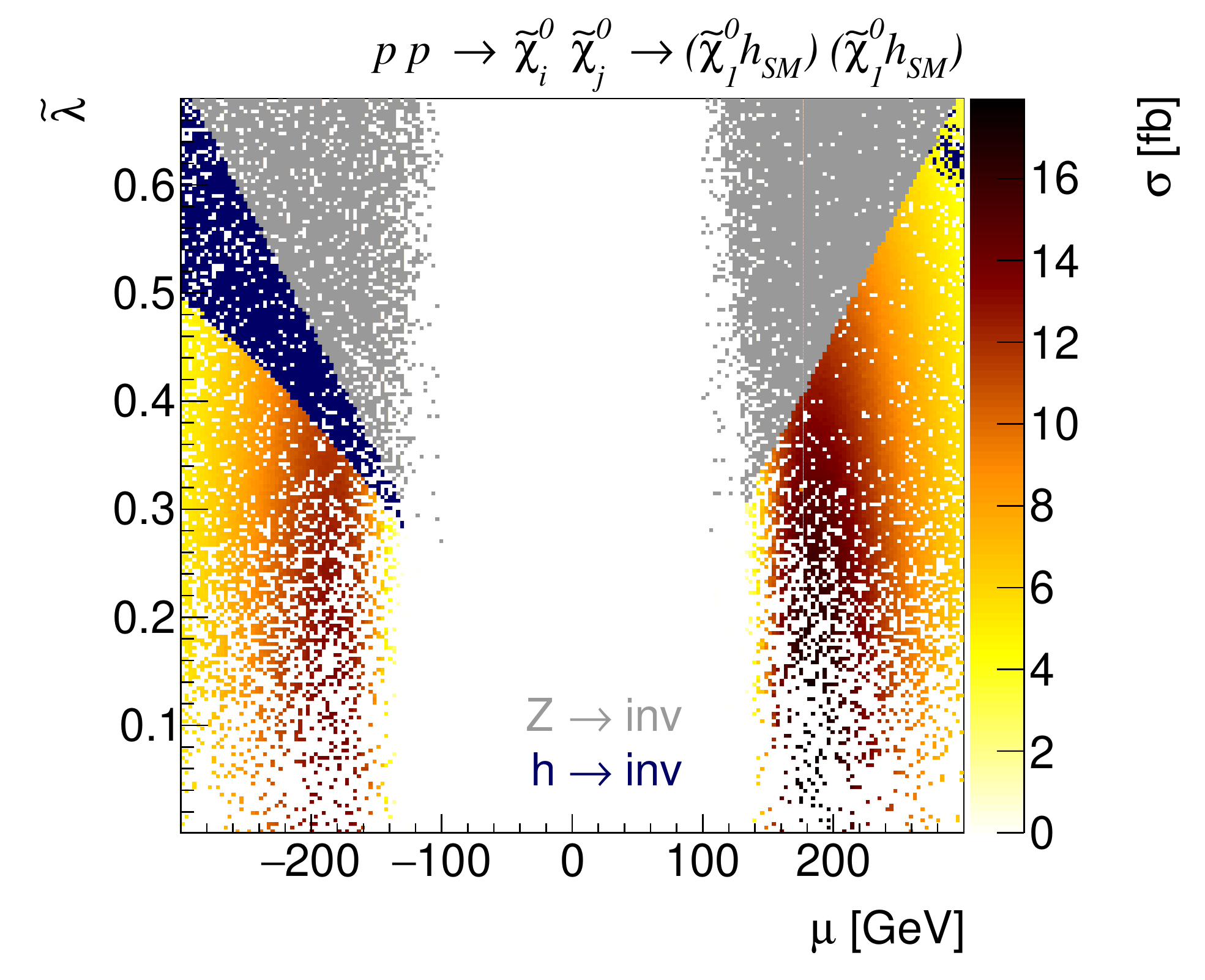}
%\hspace*{0.02\textwidth}
%\includegraphics[width=0.485\textwidth]{figures/nmssm_monohpairs_cs_muma_relicdd}
\caption{Cross section profiles for the mono-$Z$ (left) and
  mono-Higgs-pair (right) processes in the $\mu-\lambda$ plane. All
  points fulfill the chargino mass bounds. Regions excluded by
  invisible $Z$ and Higgs decays are shown in light gray and dark
  blue.}
\label{fig:nmssm_monozdihcs}
\end{figure}
%------------------------------------------------------------

While it is generally possible to extend all MSSM analyses of
Sec.~\ref{sec:fsr} to the NMSSM we focus on the two most interesting
cases, the strongly constrained mono-$Z$ signal and the most flexible
mono-Higgs-pair signal
\begin{align}
pp \to \nz{1} \nz{1} \; Z 
\qquad \text{and} \qquad 
pp \to \nz{i} \nz{j} \to ( \nz{1} h ) \; ( \nz{1} h ) \; .
\label{eq:proc_nmssm} 
\end{align}
For the mono-$Z$ signal the ISR, invisible SM-like Higgs $h_{125}$, and
heavy neutralino topologies shown in Fig.~\ref{fig:feyn_monoz_mssm}
are supplemented by the associated $Zh_s$ mediator production.

As usual, we start with the cross sections without the dark matter
constraints in Fig.~\ref{fig:nmssm_monozdihcs}. Because we fix the
dark matter mass to 10~GeV, there is no threshold left to
consider. Instead, we show the correlation between the higgsino mass
and the singlino--higgsino mixing parameter $\tilde \lambda$.  In
general, the LHC cross section grows with $\lambda$, since all
contributing diagrams are driven by bino-higgsino mixing, times
$\tilde \lambda$ connecting the higgsino content to the singlino
content.  For heavy gaugino masses, the $Z$ and Higgs decay
constraints limit the size of the higgsino fraction of the lightest
neutralino, or $\tilde \lambda$ for a given value of $\mu$.  While in
the MSSM the invisible Higgs limits were stronger for $\mu>0$, they
now constrain mostly $\mu<0$. This is because of the sign difference
between the singlino--Higgs coupling and the bino-Higgs coupling,
\begin{align}
g_{h_{125}\nz{1}\nz{1}}\approx 
\begin{cases}
g'N_{11}s_\beta\left(-\dfrac{N_{13}}{t_\beta}+N_{14}\right) \quad & \text{bino} \\
-\sqrt{2}\lambda N_{15}\left(N_{13}+\dfrac{N_{14}}{t_\beta}\right) \quad & \text{singlino.}
\end{cases}
\label{eq:singlino_coup}
\end{align}
The mono-Higgs-pairs rate shown in the right panel of
Fig.~\ref{fig:nmssm_monozdihcs} are similar to the NMSSM case shown in
Fig.~\ref{fig:monodih_cs}. As expected from the enhanced flexibility
in all couplings, they prefer a small higgsino mass and can exceed the
mono-$Z$ rates.\bigskip

%------------------------------------------------------------
\begin{figure}[t]
\includegraphics[width=0.485\textwidth]{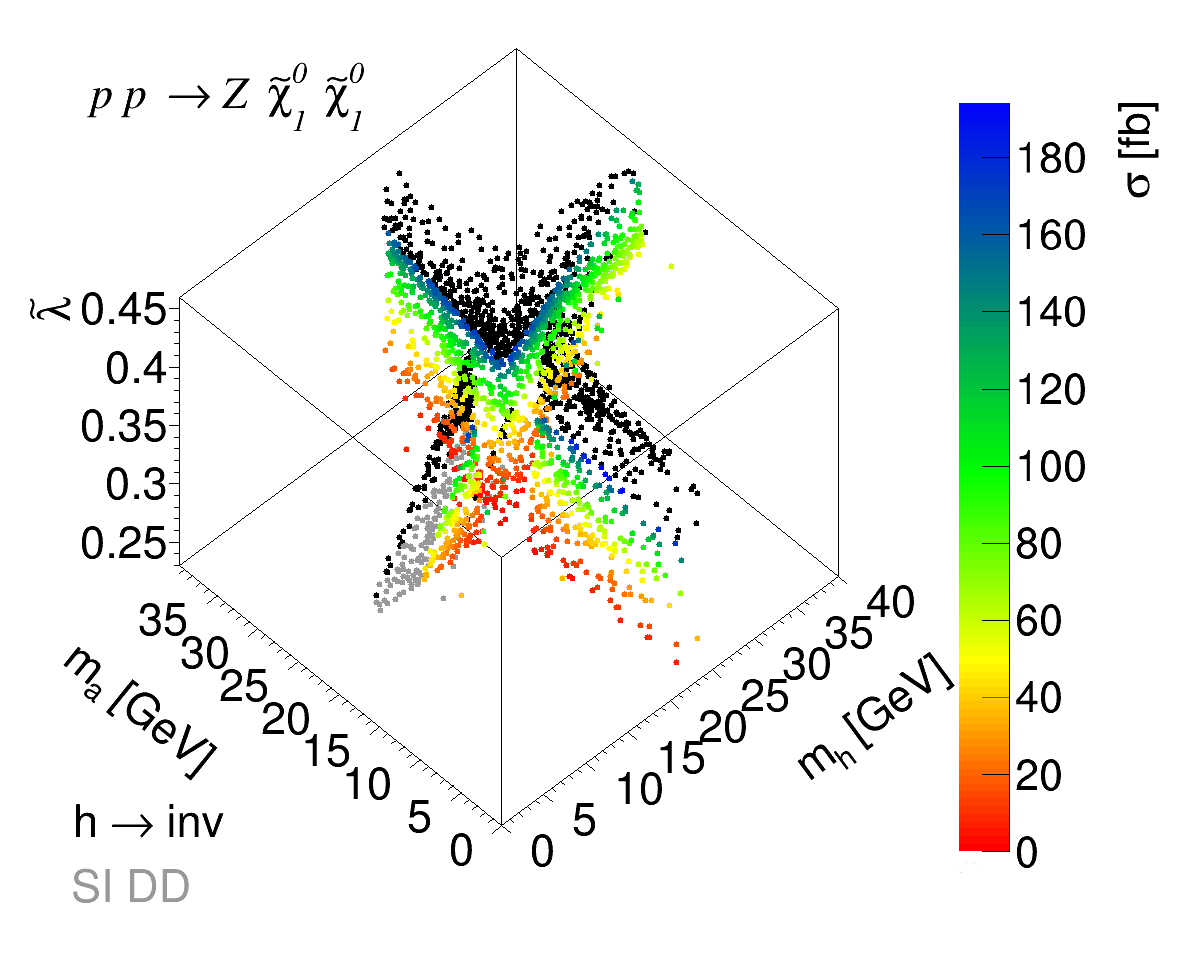}
\hspace*{0.02\textwidth}
\includegraphics[width=0.485\textwidth]{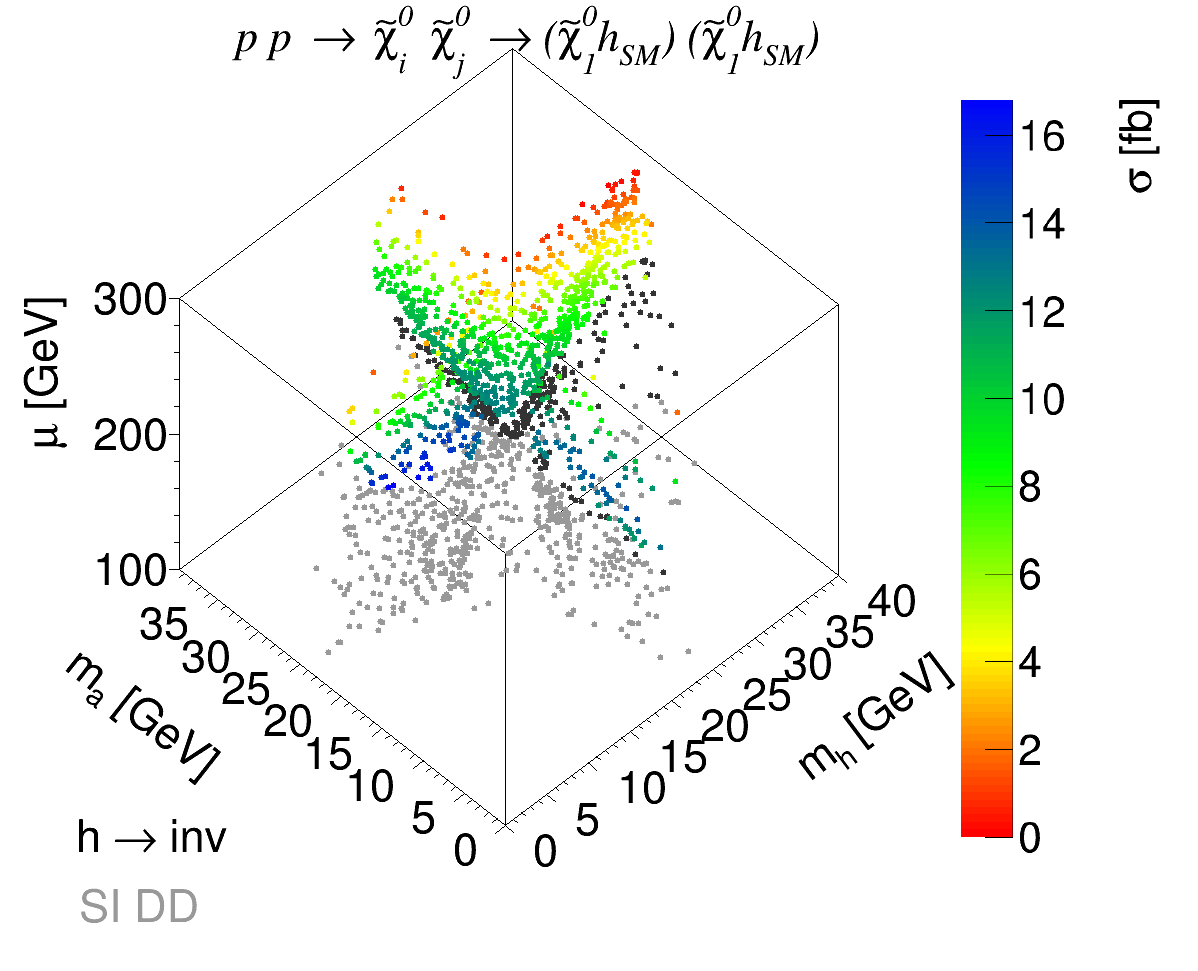}
\caption{Mono-$Z$ (left) and mono-Higgs-pair (right) cross section
  versus the singlet-like pseudoscalar and scalar masses and the
  singlino--higgsino mixing parameter.
  All points fulfill the relic density, chargino mass, and invisible
  $Z$ decay bounds. The effects of the invisible Higgs decays and
  spin-independent direct detection are shown in grey.}
\label{fig:nmssm_monozdihcs_poles}
\end{figure}
%------------------------------------------------------------

The interesting question is, how these large LHC rates change when we
apply the constraints from the relic density and direct detection.  In
the left panel of Fig.~\ref{fig:nmssm_monozdihcs_poles} we show the
results for mono-$Z$ production in the NMSSM framework. The general
pattern confirms that either the scalar or the pseudo-scalar mediator
has to be just slightly off its mass shell, with a width given by the
velocity distribution. The main difference between them arises from
CMB bounds, which are irrelevant for scalar p-wave annihilation, while
a 10~GeV neutralino is barely allowed for s-wave annihilation through
the pseudoscalar. In addition, following Eq.\eqref{eq:singlino_coup}
the LHC production rate is roughly proportional to a factor $\lambda$
from the explicit couplings and another factor $\tilde \lambda$ from
the higgsino fractions.

After including all constraints, the $Zh$ topology with an invisible
Higgs again emerges as the dominant mono-$Z$ process. However, while
for the MSSM the direct detection constraints effectively enforce
$\br\left(h_{125}\to\nz{1}\nz{1}\right) \lesssim 0.003$, they now fall
behind the LHC limit of 24\%. This way, the LHC rate in the NMSSM can
be forty times as large as in the MSSM, exceeding 100~fb.
The light, new scalar mediator also leads to spin-independent
singlino--nucleon scattering. This manifests itself in the excluded
points at low $m_{h_S}$ and large singlino--higgsino mixing
$\lambda$.\bigskip

Also in Fig.~\ref{fig:nmssm_monozdihcs_poles} we show the same effects
for mono-Higgs-pair production. In that case the relevant third
parameter is not the singlino--higgsino mixing, but the higgsino mass
parameter. In the NMSSM the LHC rates can be three times as
large as in the MSSM. The reason is a kinematic effect, because the
weaker DD bounds for smaller dark matter masses allow for a larger
higgsino fraction in the dark matter candidate and hence lighter
on-shell higgsinos.  The subsequent branching ratios for the decays
$\nz{2,3} \to h_{125} \nz{1}$ are similar to typical MSSM values,
namely around 40\%.  For constant $\mu$ this branching ratio is
approximately independent of $\lambda$, since both $g_{h\nz{1}\nz{1}}$
and $g_{Z\nz{1}\nz{1}}$ are proportional to $\lambda^2$ from the
explicit and implicit dependences.\bigskip

Altogether, we indeed see how the light NMSSM mediators allow us to
decouple the different relic density, direct detection, and LHC
observables. Most importantly, our dark matter singlet as well as the
heavier higgsinos can now be lighter than in the MSSM.  For all
channels this directly translates into an increase of the LHC rate by
a factor three to forty. The mono-$W$ channel will obviously follow the
same pattern.

%%%%%%%%%%%%%%%%%%%%%%%%%%%%%%%%%%%%%%%%%%%%%%%%%%%%%%%%%%%%
\section{Summary}
\label{sec:summary}

Mono-$X$ searches are promising strategies to search for dark matter
at the LHC. A wealth of models motivate a large number of analyses,
including mono-jet, mono-photon, mono-$Z$, mono-$W$, and mono-Higgs
searches. Major support for these searches comes from effective
theories of dark matter, but with an approach-specific ranking of the
different mono-$X$ signatures. In this paper we have compared
different mono-$X$ searches based on two orthogonal theoretical
assumptions concerning their LHC production.\medskip

As a starting point, we have compared mono-jet, mono-photon, and
mono-$Z$ searches for for a dark matter toy model with a $Z'$
mediator. In that case all mono-$X$ searches rely on the same ISR
topology and can be compared directly.  We confirmed that mono-jet
searches are by far the most promising, as long as the systematic
uncertainties are under control. Combining different ISR-based
mono-$X$ searches does not add a significant amount of information, so
we can skip ISR signatures for the remaining analysis.\medskip

In the main part of the paper we have analyzed decays of heavier
states of an extended dark matter sector as a source of electroweak
mono-$X$ signals. For electroweakinos in the MSSM we have shown how
different intermediate on-shell states lead to large predicted LHC
rates, clearly separating our topologies from any effective theory
description. Adding relic density and direct detection constraints, we
found that mono-$Z$ rates at the LHC cannot be large, if we stick to
SM-like mediators. A leading channel turned out to be associated
$Zh$ production with an invisible Higgs decay. This associated
production signature is known to be weaker than searching for the same
invisible Higgs decay in weak boson fusion.

Unlike for effective theories, we found that mono-$W$ searches are
more promising than mono-$Z$ searches, also due to a mono-$W$-pair
topology. Experimentally, we need to ensure that the contributing
$W$-pair topology is not removed by lepton or jet vetoes as part of the
mono-$W$ analysis. Initially, mono-Higgs production looks distinctly
un-promising, but once we include direct detection constraints the
combination of mono-Higgs and mono-Higgs-pair topologies allows us to
separate the LHC signal from the particles and couplings driving dark
matter annihilation and direct detection.\medskip

The NMSSM with its light scalar and pseudo-scalar mediators decouples
the relic density, direct detection and LHC rates especially for a
light dark matter. In that sense, its extended scalar sector comes
much closer to the anything-goes philosophy of simplified
models. Correspondingly, the mono-$Z$, mono-$W$(-pair), and
mono-Higgs-pair rates at the LHC can be much larger than in the
MSSM. Our two leading effects, namely that mono-$W$ production is at
least as attractive as mono-$Z$ production and that mono-$W$-pair and
mono-Higgs-pair topologies should be included, remain.

%%%%%%%%%%%%%%%%%%%%%%%%%%%%%%%%%%%%%%%%%%%%%%%%%%%%%%%%%%%%
\subsubsection*{Acknowledgments}

We would like to thank Martin Bauer for many fruitful and fun
discussions.  The authors acknowledge support by the state of
Baden-W\"urttemberg through bwHPC. TP is supported by the DFG
Forschergruppe \emph{New Physics at the LHC} (FOR~2239). 

\end{fmffile}
%%%%%%%%%%%%%%%%%%%%%%%%%%%%%%%%%%%%%%%%%%%%%%%%%%%%%%%%%%%%

\end{document}